%% file: main.tex
\documentclass[fleqn,usenatbib]{mnras}

\usepackage{newtxtext,newtxmath}

\usepackage[T1]{fontenc}
\usepackage{ae,aecompl}

\usepackage{times} 
\usepackage{tikz}
\usetikzlibrary{positioning}
\usepackage{xspace,ulem}

\usetikzlibrary{backgrounds}
\usetikzlibrary{arrows}
\usetikzlibrary{calc}
\usepackage{booktabs}
\usepackage{graphicx}	
\usepackage{amsmath}	
\usepackage{times}

\title[Cluster scaling relations in $f(R)$ gravity]{A general framework to test gravity using galaxy clusters III: \\ Observable-mass scaling relations in $f(R)$ gravity}

\author[M. A. Mitchell et al.]{
Myles A. Mitchell,$^{1}$\thanks{E-mail: m.a.mitchell@durham.ac.uk}
Christian Arnold$^{1}$
and Baojiu Li$^{1}$
\\
$^{1}$Institute for Computational Cosmology, Department of Physics, Durham University, South Road, Durham DH1 3LE, UK\\
}

\date{Accepted XXX. Received YYY; in original form ZZZ}

\pubyear{2019}

\begin{document}
\label{firstpage}
\pagerange{\pageref{firstpage}--\pageref{lastpage}}
\maketitle

\begin{abstract}
We test two methods, including one that is newly proposed in this work, for correcting for the effects of chameleon $f(R)$ gravity on the scaling relations between the galaxy cluster mass and four observable proxies. Using the first suite of cosmological simulations that simultaneously incorporate both full physics of galaxy formation and Hu-Sawicki $f(R)$ gravity, we find that these rescaling methods work with a very high accuracy for the gas temperature, the Compton $Y$-parameter of the Sunyaev-Zel'dovich (SZ) effect and the X-ray analogue of the $Y$-parameter. This allows the scaling relations in $f(R)$ gravity to be mapped to their $\Lambda$CDM counterparts to within a few percent. We confirm that a simple analytical tanh formula for the ratio between the dynamical and true masses of haloes in chameleon $f(R)$ gravity, proposed and calibrated using dark-matter-only simulations in a previous work, works equally well for haloes identified in simulations with two very different -- full-physics and non-radiative -- baryonic models. The mappings of scaling relations can be computed using this tanh formula, which depends on the halo mass, redshift and size of the background scalar field, also at a very good accuracy. Our results can be used for accurate determination of the cluster mass using SZ and X-ray observables, and will form part of a general framework for unbiased and self-consistent tests of gravity using data from present and upcoming galaxy cluster surveys. We also propose an alternative test of gravity, using the $Y_{\rm X}$-temperature relation, which does not involve mass calibration.
\end{abstract}




\section{Introduction}
\label{sec:introduction}

Clusters of galaxies are the largest gravitationally-bound objects in the Universe and, as tracers of the highest peaks of the primordial density perturbations, they are frequently used to test cosmological models. For example, global properties, such as the abundance of clusters, are expected to be highly sensitive to the values of cosmological parameters and to the strength of gravity on large scales, and they can also be predicted accurately with numerical simulations. These simulations can be combined with state-of-the-art observational data to place tight constraints on a wide range of cosmological models, including modifications to the theory of General Relativity (GR). The wealth of quality data that will be made available from ongoing and upcoming galaxy cluster surveys \citep[e.g.,][]{ukidss, desi, euclid, lsst, xmm-newton, chandra, erosita, Planck_SZ_cluster, act} has the potential to revolutionise the precision of these constraints.

Accurate measurement of the cluster mass is vital in tests that make use of probes such as cluster number counts and the gas mass fraction. However, direct estimation of the cluster mass is generally an expensive task which typically requires long exposure times, making it challenging to repeat for very large samples of clusters. A common alternative technique is to infer the cluster mass by using its relation to more easily measured observables including the X-ray luminosity and the integrated Sunyaev-Zel'dovich (SZ) flux. These mass `proxies' are intrinsically linked to the gravitational potential and, as a result, they typically have a power-law mapping with the mass. A lot of work has therefore been carried out to study and calibrate these cluster mass scaling relations using various approaches: simulations that employ a comprehensive galaxy formation model (full physics), including cooling, star formation and feedback, have been used \citep[e.g.,][]{Nagai:2007,Fabjan:2011,Truong:2016egq,2018MNRAS.480.2898C}; internal calibration has been employed via a joint likelihood analysis \citep[][]{Mantz:2010,Mantz:2015}; self-calibration can be achieved using additional observables, such as the clustering of clusters \citep[e.g.,][]{Schuecker:2003,Majumdar_2004}; and also subsamples of a complete data set can be used. Examples of the latter include cross-checking data for X-ray or SZ proxies with weak lensing data \citep[e.g.,][]{Vikhlinin:2009,Nagarajan:2018ajl}, using an existing X-ray scaling relation to calibrate an SZ scaling relation \citep[e.g.,][]{Ade:2013lmv}, and combining optical richness data with CMB-cluster lensing data \citep[e.g.,][]{Raghunathan:2018azn}. It has also been shown that characterising clusters using their gravitational potential instead of their mass can be an alternative way to reduce the cluster mass measurement bias \citep{Tchernin:2020vbt}.

A number of modified gravity (MG) theories \citep[see, e.g.,][]{Koyama:2015vza} have been proposed over the past couple of decades in an attempt to better understand the as yet unexplained accelerated cosmic expansion. In these models, the law of gravity is typically modified, which means that clusters can become more (or less) massive than their counterparts in the standard $\Lambda$ cold dark matter ($\Lambda$CDM) model. This can have consequences on the measured cluster number count, making the latter a potentially powerful cosmological probe to constrain deviations from GR.

A popular working example is the $f(R)$ gravity model \citep[see, e.g.,][]{Sotiriou:2008rp,DeFelice:2010aj}, which, like many other theories, predicts the presence of an extra scalar field that can lead to a strengthened force of gravity which is enhanced compared to Newtonian gravity. It is predicted that this could leave observational signatures in the formation of large-scale structures (LSS) that can make the model distinguishable from GR. A number of studies have therefore made use of LSS probes, such as the halo mass function probed by different proxies \citep[see, e.g.,][]{PhysRevD.92.044009,Liu:2016xes,Peirone:2016wca}, redshift-space distortions \citep[e.g.,][]{Bose:2017dtl,Hernandez-Aguayo:2018oxg,2018NatAs...2..967H}, the cluster gas mass fraction \citep[e.g.,][]{Li:2015rva}, the cluster temperature-mass relation \citep{Hammami:2016npf} and SZ profile \citep{deMartino:2016xso}, and the clustering of clusters \citep{Arnalte-Mur:2016alq} to constrain these theories. For the current status and future prospect of testing gravity using galaxy clusters, see the recent reviews by \citet{Cataneo:2018mil,Baker:2019gxo}.

\begin{figure*}
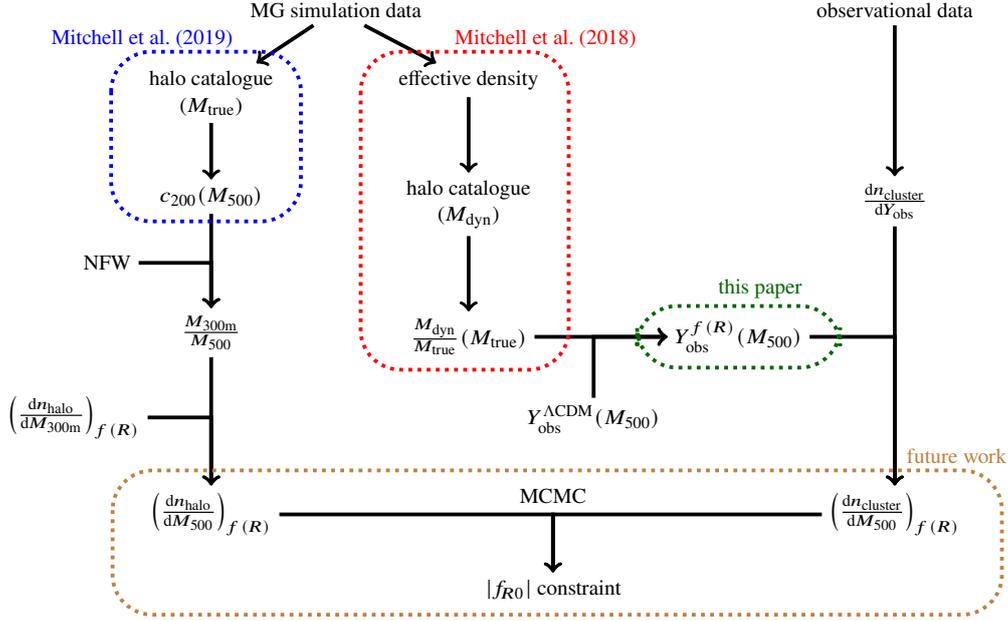

\centering
\include{flow_chart}
\caption{[{\it Colour Online}] Flow chart summarising the key steps of our framework to constrain modified gravity using galaxy clusters. For now, we focus on tests that use the halo mass function to constrain the $f(R)$ gravity model proposed by \citet{Hu:2007nk}. In previous works \citep{Mitchell:2018qrg,Mitchell:2019qke}, we used a suite of dark-matter-only simulations, which cover a wide range of mass resolutions and model parameters, to calibrate models for the enhancement of the dynamical mass (red dotted box) and the concentration (blue dotted box) of dark matter haloes in $f(R)$ gravity. The model for the concentration can be used to convert between different halo mass definitions, which can allow for the derivation of the halo mass function in $f(R)$ gravity using predictions from previous works \citep[e.g.,][]{Cataneo:2016iav}. In this work (green dotted box), we use full-physics hydrodynamical simulations to test predictions that the dynamical mass enhancement can be used to convert scaling relations calibrated in $\Lambda$CDM into their $f(R)$ gravity counterparts. These scaling relations can then be used to derive the cluster mass function, assuming $f(R)$ gravity, starting from observational data. Finally, we will use Markov Chain Monte Carlo analysis to constrain the present-day background scalar field magnitude, $|f_{R0}|$, using our theoretical predictions and observational calculations of the mass function (brown dotted box).}
\label{fig:flow_chart}
\end{figure*}

The additional forces in the MG models can have complicated effects on the internal properties of galaxy clusters, including their density profile, dynamical mass and gas temperature. However, a robust and complete pipeline to account for these effects has not yet been implemented in tests that constrain MG theories using galaxy clusters. This paper is part of a series of works that aim to propose a framework for producing unbiased and self-consistent constraints of $f(R)$ gravity (and other chameleon theories) using the halo mass function (and other LSS probes). The framework is described in Fig.~\ref{fig:flow_chart} \citep[for a more complete description, see][]{Mitchell:2018qrg}. Several parts of it are already complete. A simple yet powerful tanh fitting formula for the force enhancement was calibrated \citep{Mitchell:2018qrg} using a suite of dark-matter-only simulations. This can be used to predict the enhancement of the dynamical mass of clusters. In particular, it can be used as an effective one-parameter description of the chameleon screening mechanism. This has already enabled us to calibrate a simple model of the $f(R)$ enhancement of the halo concentration \citep{Mitchell:2019qke}, and it is expected to simplify the calibration of the $f(R)$ halo mass function in a future work. Our models for the enhancement of the dynamical mass and the concentration show a high level of accuracy for a wide and continuous range of background scalar field values, redshifts and halo masses. While these simplified and calibrated models of cluster properties were done using a particular $f(R)$ model, their underlying logic implies that they will work for generic chameleon-type MG models, promising the generality of the framework.

In this work, we analyse the effects of $f(R)$ gravity on cluster observable-mass scaling relations. These are modified by the effects of the fifth force on the gravitational potentials of haloes, which are intrinsically linked to the dynamical mass and gas temperature. According to \citet{He:2015mva}, who employed a suite of non-radiative simulations to investigate the effect of $f(R)$ gravity on a number of mass proxies, it is possible to map between the scaling relations in $f(R)$ gravity and GR using only the relation between the dynamical and true (or lensing) masses of a halo, which is accurately captured by our tanh fitting formula mentioned above (cf.~Fig.~\ref{fig:flow_chart}). Here, we test these predictions using non-radiative hydrodynamic simulations with much higher resolutions, and build upon this by checking how the addition of full-physics\footnote{Throughout this work, following the convention of the IllustrisTNG community, `full-physics' refers to the most advanced hydrodynamics scheme that is currently implemented in our cosmological simulations (see Sec.~\ref{sec:simulations}). It should not be understood as a complete description of all the physics underlying galaxy formation, much of which is far beyond the resolution limit of our simulations.} effects such as cooling, star formation and feedback impact the accuracy. To this end, we make use of the first simulations that simultaneously incorporate both full-physics and $f(R)$ gravity \citep{Arnold:2019vpg}. In addition, we propose and test a set of alternative mappings from the 
scaling relations in GR to their $f(R)$ counterparts, which again require only our tanh fitting formula. Our simulations cover galaxy groups and low-mass clusters, and we are currently unable to test the mappings for high-mass clusters ($M\gtrsim10^{14.5}M_\odot$) -- this will be left for future works that make use of larger hydrodynamics simulations of $f(R)$ gravity currently under development.

The paper is organised as follows: in Sec.~\ref{sec:background}, several important aspects of the $f(R)$ gravity theory are introduced, and we summarise the scaling relation mappings proposed by \citet{He:2015mva} and our novel alternative mappings; in Sec.~\ref{sec:methods}, we describe our simulations and calculations of the halo masses and observable proxies; then, in Sec.~\ref{sec:results}, we present our results for the scaling relations of four mass proxies; finally, in Sec.~\ref{sec:conclusions}, we summarise the results of this work and their significance for our framework.

\section{Background}
\label{sec:background}

We describe several key aspects of the \citet{Hu:2007nk} $f(R)$ gravity model in Sec.~\ref{sec:background:theory}. Then, in Sec.~\ref{sec:background:scaling_relations}, we outline the effect of $f(R)$ gravity on the dynamical mass of dark matter haloes and summarise predictions for how several cluster mass scaling relations are affected. We discuss the mappings between the $f(R)$ and GR mass scaling relations proposed by \citet{He:2015mva} (Sec.~\ref{sec:eff_approach}) and outline a new alternative approach (Sec.~\ref{sec:true_approach}) which is also tested in this work.

\subsection{\boldmath The $\lowercase{f}(R)$ gravity model}
\label{sec:background:theory}

The $f(R)$ gravity theory represents a modification to GR. The modification takes the form of the addition of a non-linear function, $f(R)$, of the Ricci scalar curvature, $R$, to the Einstein-Hilbert action $S$:
\begin{equation}
    S=\int {\rm d}^4x\sqrt{-g}\left[\frac{R+f(R)}{16\pi G}+\mathcal{L}_{\rm M}\right],
\label{eq:action}
\end{equation}
where $g$ is the determinant of the metric tensor, $\mathcal{L}_{\rm M}$ the Lagrangian density for matter and $G$ the (universal) gravitational constant. This alteration leads to the introduction of an extra tensor, $X_{\alpha \beta}$, in the Einstein field equations:
\begin{equation}
    G_{\alpha \beta} + X_{\alpha \beta} = 8\pi GT_{\alpha \beta},
\label{eq:modified_field_equations}
\end{equation}
where
\begin{equation}
    X_{\alpha \beta} = f_RR_{\alpha \beta} - \left(\frac{f}{2}-\Box f_R\right)g_{\alpha \beta} - \nabla_{\alpha}\nabla_{\beta}f_R.
\label{eq:GR_modification}
\end{equation}
Here the tensors $G_{\alpha \beta}$, $T_{\alpha \beta}$ and $R_{\alpha \beta}$ represent the Einstein tensor, the stress-energy tensor and the Ricci curvature, respectively. The d'Alembert operator is represented by $\Box$, and $\nabla_{\alpha}$ is the covariant derivative associated with $g_{\alpha \beta}$, the metric tensor. The scalar field $f_R\equiv{\rm d}f(R)/{\rm d}R$ represents the extra (scalar) degree of freedom of the theory. This mediates a so-called `fifth force', which is an attractive force felt by massive particles. When this force is able to act, the effect is to enhance the strength of gravity by a factor of $4/3$. The force can act on scales smaller than the Compton wavelength, $\lambda_{\rm C}$, which sets the physical range of it,
\begin{equation}
    \lambda_{\rm C} = a^{-1}\left(3\frac{{\rm d}f_R}{{\rm d}R}\right)^{\frac{1}{2}},
\label{eq:compton_wavelength}
\end{equation}
where $a$ is the cosmological scale factor. Beyond $\lambda_{\rm C}$ the amplitude of this force decays exponentially.

Solar system tests \citep[e.g.,][]{Will:2014kxa} confirm GR to a very high precision. For consistency with these tests, the chameleon screening mechanism is employed in the $f(R)$ gravity model \citep[e.g.,][]{Khoury:2003aq,Khoury:2003rn}. It introduces an environment-dependent effective mass of the scalar field $f_R$, which becomes more massive in high-density regimes. This has the effect of decreasing $\lambda_{\rm C}$ and thus screening out the fifth force in high-density regions, including the solar system.

Following previous works, we focus on a representative variant of $f(R)$ gravity proposed by \citet[][HS]{Hu:2007nk}. This model is able to give rise to the late-time cosmic acceleration, while also showing consistency with solar system tests. It is characterised by the following prescription for $f(R)$:
\begin{equation}
    f(R) = -m^2\frac{c_1\left(-R/m^2\right)^n}{c_2\left(-R/m^2\right)^n+1},
\label{eq:hu_sawicki}
\end{equation}
where $m^2\equiv8\pi G\bar{\rho}_{\rm M,0}/3=H_0^2\Omega_{\rm M}$, with $H_0$ the Hubble constant, $\bar{\rho}_{\rm M,0}$ the present-day background matter density and $\Omega_{\rm M}$ the current dimensionless matter density parameter. The model has three parameters: $c_1$, $c_2$ and $n$. However, for a realistic set of cosmological parameters, it can be re-formulated to a good accuracy with just two free parameters: $n$, which is fixed at $1$ in this work, and the present-day background scalar field value $f_{R0}$ (we note that $f_R(z)$ and $f_{R0}$ represent background values for the remainder of this paper). We investigate models with values $|f_{R0}|=10^{-6}$ and $|f_{R0}|=10^{-5}$, referring to these as F6 and F5, respectively. F5 represents a stronger modification to GR than F6, and allows higher-mass haloes to be unscreened.

\subsection{\boldmath Mass scaling relations in $\lowercase{f}(R)$ gravity}
\label{sec:background:scaling_relations}

Various techniques can be employed to measure the mass of clusters. One approach is to use the gravitational lensing of background sources. In $f(R)$ gravity, apart from an overall rescaling of the lens mass by $\left(1+f_R\right)^{-1}$, which is very close to unity for realistic choices of the $f(R)$ parameters, photons are unaffected by the fifth force, meaning that the lensing effect of a given cluster is the same as in GR. In this work, we refer to the mass inferred from lensing as the `true' mass, $M_{\rm true}$, and this satisfies the following: $M^{f(R)}_{\rm true}=M^{\rm GR}_{\rm true}$. The mass can alternatively be measured using probes such as the X-ray temperature. These can provide a measure of the `dynamical' mass, $M_{\rm dyn}$, which we define as the mass that is felt by nearby massive test particles. Because the gravitational potential of unscreened haloes is enhanced in $f(R)$ gravity, the dynamical mass of haloes is also enhanced by the same factor: $M^{f(R)}_{\rm true}\leq M^{f(R)}_{\rm dyn}\leq(4/3)M^{f(R)}_{\rm true}$. On the other hand, in GR, the dynamical mass is expected to be equal to the true mass: $M^{\rm GR}_{\rm dyn}=M^{\rm GR}_{\rm true}=M^{\rm GR}$. 

Using a suite of $N$-body simulations covering a wide range of resolutions, \citet{Mitchell:2018qrg} showed that the enhancement of the dynamical mass in $f(R)$ gravity can be accurately described by a simple $\tanh$ formula, given by,
\begin{equation}
\frac{M^{f(R)}_{\rm dyn}}{M^{f(R)}_{\rm true}} = \frac{7}{6}-\frac{1}{6}\tanh\left(p_1\left[\log_{10}\left(M^{f(R)}_{\rm true}M_{\odot}^{-1}h\right)-p_2\right]\right).
\label{eq:enhancement}
\end{equation}
The parameter $p_1$ was found to be approximately constant with value $2.21\pm0.01$, while a simple power law, whose slope was motivated by theory (which predicts it to be $3/2$), was calibrated for $p_2$:
\begin{equation}
    p_2=(1.503\pm0.006)\log_{10}\left(\frac{|f_R(z)|}{1+z}\right)+(21.64\pm0.03).
\label{eq:p2}
\end{equation}
This depends on just a single parameter, $|f_R|/(1+z)$, and provides a powerful yet simple approach to describe the chameleon screening mechanism in $f(R)$ gravity. 

In observational studies, obtaining detailed and high-quality X-ray and spectral data for determination of the cluster dynamical mass is often an expensive task, requiring long exposure times. It is therefore more common to predict the cluster mass using its relationship with more easily measured observables, including the temperature, the X-ray luminosity, and the Compton $Y$-parameter of the SZ effect, and its X-ray analogue. These mass `proxies' have a one-to-one mapping with the mass because of the link between the gravitational potential of a cluster and its temperature. During cluster formation, baryonic matter is accreted onto the dark matter halo from its surroundings. The gravitational potential energy of the gas is converted into kinetic energy as it falls in. During accretion, the in-falling gas undergoes shock heating, resulting in the conversion of its kinetic energy into thermal energy. The resulting self-similar model for cluster mass scaling relations predicts that the gravitational potential alone can determine the thermodynamical properties of a cluster \citep{1986MNRAS.222..323K}.

In this work, we study the effects of HS $f(R)$ gravity on the scaling relations for four mass proxies: the gas temperature $T_{\rm gas}$, the X-ray luminosity $L_{\rm X}$, the integrated SZ flux, given by the Compton $Y$-parameter $Y_{\rm SZ}$, and its X-ray analogue $Y_{\rm X}$. The X-ray luminosity within radius $r$ from the cluster centre is given by,
\begin{equation}
    L_{\rm X}(<r) = \int^r_0{\rm d}r'4\pi r'^2\rho_{\rm gas}^2(r')T_{\rm gas}^{1/2}(r'),
    \label{eq:lx_obs}
\end{equation}
where $\rho_{\rm gas}(r)$ is the gas density profile. The $Y_{\rm SZ}$ parameter is related to the integrated electron pressure of the cluster gas, and is given by,
\begin{equation}
     Y_{\rm SZ}(<r) = \frac{\sigma_{\rm T}}{m_{\rm e}c^2}\int^r_0{\rm d}r'4\pi r'^2n_{\rm e}(r')T_{\rm gas}(r'),
     \label{eq:ysz_obs}
\end{equation}
where $\sigma_{\rm T}$ is the Thomson scattering cross section, $m_{\rm e}$ is the electron rest mass, $c$ is the speed of light and $n_{\rm e}$ is the electron number density. Meanwhile, the $Y_{\rm X}$ parameter \citep{Kravtsov:2006db} is equivalent to the product of the gas mass and the mass-weighted gas temperature, $\bar{T}_{\rm gas}(<r)$:
\begin{equation}
    Y_{\rm X}(<r) = \bar{T}_{\rm gas}(<r)\int^r_0{\rm d}r'4\pi r'^2\rho_{\rm gas}(r').
    \label{eq:yx_obs}
\end{equation}

It has been shown in previous studies \citep[e.g.,][]{Fabjan:2011} that $Y_{\rm X}$ and $Y_{\rm SZ}$ have comparatively low scatter as mass proxies and are relatively insensitive to dynamical processes including cluster mergers. It has also been found that their scaling relations with the mass show good agreement with the self-similar model predictions even after the inclusion of full-physics effects such as feedbacks, which can heat up and blow gas out from the central regions \citep[e.g.,][]{Fabjan:2011,Truong:2016egq,2018MNRAS.480.2898C}. 

Because these cluster mass proxies are closely related to the dynamical mass of clusters, and because in $f(R)$ gravity the dynamical mass can be cleanly modelled (see above), it is natural to expect that cluster observable-mass scaling relations in $f(R)$ gravity can be modelled given their counterparts in GR. To study the effects of $f(R)$ gravity on the scaling relations for these proxies, we adopt two methods which are described in the sections that follow.

\subsubsection{Effective density approach}
\label{sec:eff_approach}

The $f(R)$ gravity effective density field, $\rho_{\rm eff}$, was originally defined by \citet{He:2014eva}:
\begin{equation}
    \delta\rho_{\rm eff} \equiv \left(\frac{4}{3}-\frac{\delta R}{24\pi G\delta\rho}\right)\delta\rho,
\end{equation}
where $\rho$ is the true density field, corresponding to the intrinsic mass of simulation particles. Written in terms of the effective density, the Poisson equation of $f(R)$ gravity is cast into the same form as in Newtonian gravity:
\begin{equation}
    \nabla^2\phi = 4\pi G\delta\rho_{\rm eff},
\end{equation}
where $\phi$ is the total gravitational potential, including contributions from the fifth force. It follows that the mass of haloes computed using the effective density field is equivalent to the dynamical mass. 

Using non-radiative simulations run for the F5 model and GR, \citet{He:2015mva} generated halo catalogues using the effective density field. The radius, $R_{\rm 500}^{\rm eff}$, of these haloes enclosed an average \textit{effective} density of 500 times the ({\it true}) critical density of the Universe. Both the total true and dynamical mass were computed within this radius, and the cluster observables were computed using all enclosed gas particles.

Analysing these data, \citet{He:2015mva} found that haloes in GR and $f(R)$ gravity with the same dynamical mass, $M^{\rm GR}=M_{\rm dyn}^{f(R)}=M_{\rm dyn}$, also have the same gas temperature:
\begin{equation}
    T^{f(R)}_{\rm gas}\left(M^{f(R)}_{\rm dyn}\right) = T^{\rm GR}_{\rm gas}\left(M^{\rm GR}=M^{f(R)}_{\rm dyn}\right).
    \label{eq:temp_equiv_eff}
\end{equation}
The physical origin of this result is the intrinsic relationship between the gravitational potential and the gas temperature (see above). Two haloes with the same dynamical mass $M_{\rm dyn}$ (which we recall has also been computed within the same radius $R_{500}^{\rm eff}$), would also have the same gravitational potential, $\phi=(GM_{\rm dyn})/R_{500}^{\rm eff}$, and are therefore expected to have the same temperature. The authors also found that, outside the core region, the gas density profiles of haloes in GR are enhanced by a factor $M^{f(R)}_{\rm dyn}/M^{f(R)}_{\rm true}$ with respect to haloes in $f(R)$ gravity which have the same dynamical mass. This is because the gas density profile follows the true density profile more closely than the effective density profile, which itself is a result of the fact that clusters form from very large regions in the Lagrangian space, so that the ratio between the baryonic and total masses within clusters resembles the cosmic mean, $\Omega_{\rm B}/\Omega_{\rm M}$ \citep{1993Natur.366..429W}, in which $\Omega_{\rm B}$ is the present-day baryonic density parameter. The extra forces in MG theories and feedbacks from galaxy formation can add further complications to this through their effects on the gas density profiles, especially in the inner regions; however, as we will show in the following paragraph, the good agreement between the GR and rescaled $f(R)$ gas density profiles still holds in the outer halo regions.

\begin{figure*}
\centering
\includegraphics[width=1.0\textwidth]{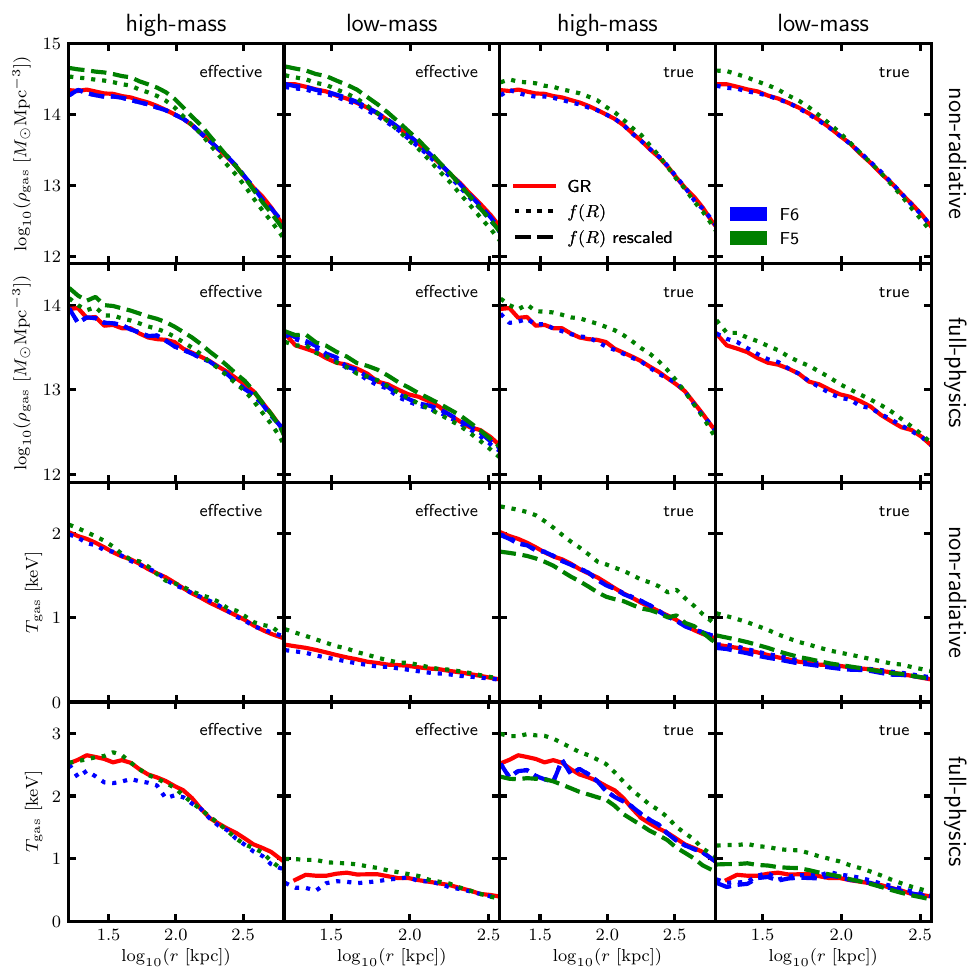}
\caption{[{\it Colour Online}] Median gas density profiles (\textit{top two rows}) and median temperature profiles (\textit{bottom two rows}) of FOF groups from two mass bins: $13.7<\log_{10}\left(M/M_{\odot}\right)<14.0$ (\textit{high-mass}) and $13.0<\log_{10}\left(M/M_{\odot}\right)<13.3$ (\textit{low-mass}). The group data from the non-radiative and full-physics SHYBONE simulations (see Sec.~\ref{sec:simulations}) has been used. In addition to GR (\textit{red solid lines}), the profiles for F6 (\textit{blue lines}) and F5 (\textit{green lines}) are shown. Rescaled $f(R)$ gravity profiles (\textit{dashed lines}) are shown, along with the unaltered profiles (\textit{dotted curves}). The rescalings correspond to the effective density (\textit{left two columns}) and true density (\textit{right two columns}) approaches discussed in Sec.~\ref{sec:background:scaling_relations}. For the effective approach, the maximum radius shown is $R_{500}^{\rm eff}$ (see Sec.~\ref{sec:eff_approach}) and the halo mass $M$ is the total dynamical mass within this radius. For the true approach, the maximum radius shown is $R_{500}^{\rm true}$ (see Sec.~\ref{sec:true_approach}) and the halo mass $M$ is the total true mass within this radius.}
\label{fig:profiles}
\end{figure*}

We have replicated the procedure adopted by \citet{He:2015mva} using our full-physics and non-radiative simulations (for full details, see Sec.~\ref{sec:methods}). In Fig.~\ref{fig:profiles}, the stacked temperature and gas density profiles of haloes from mass bins $10^{13.7}M_{\odot}<M_{\rm dyn}(<R_{500}^{\rm eff})<10^{14.0}M_{\odot}$ and $10^{13.0}M_{\odot}<M_{\rm dyn}(<R_{500}^{\rm eff})<10^{13.3}M_{\odot}$ are shown in the first and second columns from the left, respectively. The radial range is shown up to the mean logarithm of $R_{500}^{\rm eff}$ (which is almost exactly the same for GR, F6 and F5). For the non-radiative temperature profiles, shown in the third row, it is clear that the F6 and F5 predictions agree very well with GR in the outer regions. There is also encouraging agreement for the full-physics data, although there is a small disparity between F5 and GR in the outer regions for the higher-mass bin. For the $f(R)$ gravity gas density profiles, the results both with and without the $M^{f(R)}_{\rm dyn}/M^{f(R)}_{\rm true}$ rescaling are shown. As was found by \citet{He:2015mva}, the rescaled $f(R)$ gravity profiles (shown by the dashed curves) agree very well with GR in the outer regions. Again, there is also promising agreement for the full-physics profiles.

These results suggest that for haloes in $f(R)$ gravity and GR which have the same dynamical mass, $M^{\rm GR}=M^{f(R)}_{\rm dyn}$, the following relation is expected to apply:
\begin{equation}
\begin{aligned}
& \int_0^r {\rm d} r'4\pi r'^2\left(\rho_{\rm gas}^{f(R)}\right)^a\left(T_{\rm gas}^{f(R)}\right)^b \\
& \approx \left(\frac{M_{\rm true}^{f(R)}}{M_{\rm dyn}^{f(R)}}\right)^a\int_0^r {\rm d} r'4\pi r'^2 \left(\rho_{\rm gas}^{\rm GR}\right)^a\left(T_{\rm gas}^{\rm GR}\right)^b,
\end{aligned}
\label{eq:eff_rescaling}
\end{equation}
where $a$ and $b$ represent indices of power, and we note that $\rho_{\rm gas}$ represents the intrinsic (not effective) gas density. Using Eqs.~(\ref{eq:lx_obs})-(\ref{eq:yx_obs}) for the mass proxies, this relation can be applied to derive the following mappings between the respective mass scaling relations in GR and $f(R)$ gravity:
\begin{equation}
    \frac{M_{\rm dyn}^{f(R)}}{M_{\rm true}^{f(R)}}Y_{\rm SZ}^{f(R)}\left(M_{\rm dyn}^{f(R)}\right) \approx Y_{\rm SZ}^{\rm GR}\left(M^{\rm GR}=M_{\rm dyn}^{f(R)}\right),
    \label{eq:ysz_mapping}
\end{equation}
\begin{equation}
    \frac{M_{\rm dyn}^{f(R)}}{M_{\rm true}^{f(R)}}Y_{\rm X}^{f(R)}\left(M_{\rm dyn}^{f(R)}\right) \approx Y_{\rm X}^{\rm GR}\left(M^{\rm GR}=M_{\rm dyn}^{f(R)}\right),
    \label{eq:yx_mapping}
\end{equation}
\begin{equation}
    \left(\frac{M_{\rm dyn}^{f(R)}}{M_{\rm true}^{f(R)}}\right)^2L_{\rm X}^{f(R)}\left(M_{\rm dyn}^{f(R)}\right) \approx L_{\rm X}^{\rm GR}\left(M^{\rm GR}=M_{\rm dyn}^{f(R)}\right).
    \label{eq:lx_mapping}
\end{equation}
Note that to obtain these relations, the two integrations in Eq.~\eqref{eq:eff_rescaling} have used the same upper limit, $r=R^{\rm eff}_{500}$, for GR and $f(R)$ gravity, as mentioned above.

\citet{He:2015mva} demonstrated an accuracy $\approx3\%$ for the $Y_{\rm SZ}$ and $Y_{\rm X}$ relations and $\approx13\%$ for $L_{\rm X}$. These quantities are all cumulative: they are computed by summing over the entire volume of a halo up to some maximum radius (in this case $R_{500}^{\rm eff}$). Cumulative quantities typically depend more on the outer regions, which contain most of the volume and mass, than on the inner regions. Therefore, even though the $f(R)$ gravity profiles (with appropriate rescaling) do not agree with GR for the inner regions (see Fig.~\ref{fig:profiles}), this is expected to have a negligible contribution overall to these mass proxies and explains why \citet{He:2015mva} found such a high accuracy for these mappings. 

In our work, we will expand on these tests by using full-physics hydrodynamical simulations to check how the addition of effects such as cooling and feedback can alter the accuracy of the mappings defined by Eqs.~(\ref{eq:ysz_mapping})-(\ref{eq:lx_mapping}) and the temperature equivalence given by Eq.~(\ref{eq:temp_equiv_eff}). Our tests with the non-radiative runs can also be used as a check for consistency with \citet{He:2015mva}, who used a different simulation code and $f(R)$ gravity solver.

\subsubsection{True density approach}
\label{sec:true_approach}

Mappings can also be done for haloes identified with the true density field. For these haloes, the radius, $R_{\rm 500}^{\rm true}$, would enclose an average \textit{true} density of 500 times the critical density of the Universe. For haloes in $f(R)$ gravity and GR with the same true mass, $M^{\rm GR} = M_{\rm true}^{f(R)}$, the total gravitational potential at $R_{\rm 500}^{\rm true}$ in $f(R)$ gravity would be a factor of $M_{\rm dyn}^{f(R)}/M_{\rm true}^{f(R)}$ higher than in GR (where both the dynamical and true mass are measured within $R_{\rm 500}^{\rm true}$). According to the self-similar model predictions, the gas temperature in $f(R)$ gravity is expected to be higher by the same factor:
\begin{equation}
    T^{f(R)}_{\rm gas}\left(M^{f(R)}_{\rm true}\right) = \frac{M_{\rm dyn}^{f(R)}}{M_{\rm true}^{f(R)}}T^{\rm GR}_{\rm gas}\left(M^{\rm GR}=M^{f(R)}_{\rm true}\right).
    \label{eq:temp_equiv_true}
\end{equation}
On the other hand, for haloes in $f(R)$ gravity and GR with the same true mass, the gas density profiles are expected to agree in the outer regions.

To check these assumptions, let us once again examine Fig.~\ref{fig:profiles}, this time looking at the third and fourth columns from the left, which show the stacked gas density and temperature profiles for groups in mass bins $10^{13.7}M_{\odot}<M_{\rm true}\left(<R_{500}^{\rm true}\right)<10^{14.0}M_{\odot}$ and $10^{13.0}M_{\odot}<M_{\rm true}\left(<R_{500}^{\rm true}\right)<10^{13.3}M_{\odot}$, respectively. The radial range is shown up to the mean logarithm of $R_{500}^{\rm true}$ for all profiles. Referring to the bottom two rows, which show the non-radiative and full-physics temperature profiles, it appears that the $f(R)$ gravity profiles with the $M_{\rm dyn}^{f(R)}/M_{\rm true}^{f(R)}$ rescaling (shown by the dashed curves) show reasonable agreement with GR in the outer regions. Again, the only exception is for the high-mass bin of the full-physics data, where there is a small disparity between F5 and GR. Looking at the top two rows, the $f(R)$ gravity gas density profiles agree very well with GR in the outer-most regions for both mass bins.

These results for the temperature and gas density profiles yield the following predictions for haloes in GR and $f(R)$ gravity with $M^{\rm GR}=M_{\rm true}^{f(R)}$:
\begin{equation}
\begin{aligned}
& \int_0^r {\rm d} r'4\pi r'^2\left(\rho_{\rm gas}^{f(R)}\right)^a\left(T_{\rm gas}^{f(R)}\right)^b \\
& \approx \left(\frac{M_{\rm dyn}^{f(R)}}{M_{\rm true}^{f(R)}}\right)^b\int_0^r {\rm d} r'4\pi r'^2 \left(\rho_{\rm gas}^{\rm GR}\right)^a\left(T_{\rm gas}^{\rm GR}\right)^b,
\end{aligned}
\label{eq:true_rescaling}
\end{equation}
where this time the two integrations both have upper limit $r=R^{\rm true}_{500}$. This prediction yields the following new mappings between the mass scaling relations in $f(R)$ gravity and GR:
\begin{equation}
    Y_{\rm SZ}^{f(R)}\left(M_{\rm true}^{f(R)}\right) \approx \frac{M_{\rm dyn}^{f(R)}}{M_{\rm true}^{f(R)}}Y_{\rm SZ}^{\rm GR}\left(M^{\rm GR}=M_{\rm true}^{f(R)}\right),
    \label{eq:ysz_mapping_true}
\end{equation}
\begin{equation}
    Y_{\rm X}^{f(R)}\left(M_{\rm true}^{f(R)}\right) \approx \frac{M_{\rm dyn}^{f(R)}}{M_{\rm true}^{f(R)}}Y_{\rm X}^{\rm GR}\left(M^{\rm GR}=M_{\rm true}^{f(R)}\right),
    \label{eq:yx_mapping_true}
\end{equation}
\begin{equation}
    L_{\rm X}^{f(R)}\left(M_{\rm true}^{f(R)}\right) \approx \left(\frac{M_{\rm dyn}^{f(R)}}{M_{\rm true}^{f(R)}}\right)^{1/2}L_{\rm X}^{\rm GR}\left(M^{\rm GR}=M_{\rm true}^{f(R)}\right).
    \label{eq:lx_mapping_true}
\end{equation}
For the $Y_{\rm SZ}$ and $Y_{\rm X}$ mappings, the $M_{\rm dyn}^{f(R)}/M_{\rm true}^{f(R)}$ factor comes from the dependence on the gas temperature to power one in Eqs.~(\ref{eq:ysz_obs}) and (\ref{eq:yx_obs}). On the other hand, the X-ray luminosity, given by Eq.~(\ref{eq:lx_obs}), depends on the gas temperature to power half, which means that the corresponding $f(R)$ gravity and GR scaling relations are expected to differ by factor $\left(M_{\rm dyn}^{f(R)}/M_{\rm true}^{f(R)}\right)^{1/2}$ only. In Sec.~\ref{sec:results}, we show the results of our tests of these alternative predictions using both our non-radiative and full-physics simulations.

In this section, we have referred to two different definitions of the halo radius: $R_{500}^{\rm eff}$ and $R_{500}^{\rm true}$. The radius $R_{500}^{\rm eff}$ is defined in terms of the effective density field. For an unscreened halo in $f(R)$ gravity, the effective density is up to $4/3$ times greater than the true density. As such, $R_{500}^{\rm eff}$ is typically a higher radius than $R_{500}^{\rm true}$ for haloes in $f(R)$ gravity. On the other hand, the two radii are exactly the same in GR, where the effective density field is equivalent to the true one.

\section{Simulations and methods}
\label{sec:methods}

In Sec.~\ref{sec:simulations}, we describe the non-radiative and full-physics simulations that are used in this work. Then, in Sec.~\ref{sec:methods:groups}, we describe how we have measured the cluster mass and four observable mass proxies from these simulations.

\subsection{Simulations}
\label{sec:simulations}

The results discussed in this work were generated using a subset of the SHYBONE simulations \citep{Arnold:2019vpg}. These full-physics hydrodynamical simulations employ the IllustrisTNG galaxy formation model \citep{2017MNRAS.465.3291W,Pillepich:2017jle} and include runs for both GR and HS $f(R)$ gravity. The simulations have been run using the \textsc{arepo} code \citep{2010MNRAS.401..791S}, which is a highly parallel and optimised code for hydrodynamical cosmological simulations. The code features a modified gravity solver which uses adaptive mesh refinement to accurately measure the fifth force in high-density environments. For every full-physics run used in this work, we also utilise a non-radiative counterpart which does not include cooling, star formation or stellar and black hole feedback processes. 

Both the full-physics and non-radiative simulations span a comoving box of length 62$h^{-1}{\rm Mpc}$. These runs each start with $512^3$ dark matter particles and the same number of initial gas resolution elements (Voronoi cells), and begin at redshift $z=127$. All results in this work are computed at $z=0$. The cosmological parameters have values ($h$, $\Omega_{\rm M}$, $\Omega_{\rm B}$, $\Omega_{\Lambda}$, $n_{\rm s}$, $\sigma_8$) $=$ ($0.6774$, $0.3089$, $0.0486$, $0.6911$, $0.9667$, $0.8159$), where $h=H_0/(100~{\rm km/s/Mpc})$, $\Omega_\Lambda=1.0-\Omega_{\rm M}$, $n_{\rm s}$ is the power-law index of the primordial density power spectrum and $\sigma_8$ is the root-mean-squared linear matter density fluctuations at $z=0$. The mass resolution is set by the DM particle mass $m_{\rm DM}=1.28\times10^8h^{-1}M_{\odot}$ and an average gas cell mass of $m_{\rm gas}\approx2.5\times10^7h^{-1}M_{\odot}$. In addition to GR, the runs include the F6 and F5 HS models, all starting from identical initial conditions at $z=127$.

In the calculation of the gas temperature, we have assumed that the primordial hydrogen mass fraction has a value $X_{\rm H}=0.76$ and set the adiabatic index to $\gamma=5/3$ (for a monatomic gas). For the non-radiative simulations we assume that the gas is made up of fully ionised hydrogen and helium.

\subsection{Group catalogues}
\label{sec:methods:groups}

The group catalogues were constructed using the \textsc{subfind} code implemented in AREPO, which employs a standard friends-of-friends (FOF) algorithm combined with an un-binding method to identify the bound structures within a FOF group and the gravitational potential minimum of the objects \citep{springel2001} using the true density field. For each group, both radii $R_{500}^{\rm true}$ and $R_{500}^{\rm eff}$ were computed around the gravitational potential minimum, enclosing, respectively, average \textit{true} and \textit{effective} densities of 500 times the critical density of the Universe. For each radius definition, the total enclosed dynamical and true masses were measured, in addition to the group observables. Quantities measured within $R_{500}^{\rm eff}$ have been used to test the scaling relation mappings from the effective density approach described in Sec.~\ref{sec:eff_approach}, while the quantities measured within $R_{500}^{\rm true}$ have been used to test the predictions of the true density approach discussed in Sec.~\ref{sec:true_approach}.

In the computation of the halo temperature, we have excluded the core regions in which the complex thermal and dynamical processes during cluster formation and evolution can lead to a significant degree of dispersion between the halo temperature profiles. We have set the core region to the radial range $r<0.15R$, where $R$ can be either $R_{500}^{\rm eff}$ or $R_{500}^{\rm true}$. This range is consistent with previous studies of cluster scaling relations \citep[e.g.,][]{Fabjan:2011,Brun:2016jtk,Truong:2016egq}.

The halo gas temperature has been computed using a mass-weighted average:
\begin{equation}
    \bar{T}_{\rm gas} = \frac{\sum_i m_{{\rm gas},i}T_i}{\sum_i m_{{\rm gas},i}},
    \label{eq:mass_weighted_temperature}
\end{equation}
where $m_{{\rm gas},i}$ and $T_i$ are, respectively, the mass and temperature of gas cell $i$. The summations have been performed over all gas cells whose positions fall within the radial range $0.15R<r<R$. The integrated SZ flux is given by:
\begin{equation}
    Y_{\rm SZ} = \frac{\sigma_{\rm T}}{m_{\rm e}c^2}\sum_i N_{{\rm e},i}T_i,
    \label{eq:ysz}
\end{equation}
where $N_{{\rm e},i}$ is the number of electrons in gas cell $i$ and the sum includes the same cells as for $\bar{T}_{\rm gas}$. The X-ray analogue of the integrated SZ flux is equal to the product of the total gas mass $M_{\rm gas}$, of all gas cells within $R_{500}$, and $\bar{T}_{\rm gas}$:
\begin{equation}
    Y_{\rm X} = M_{\rm gas}\times \bar{T}_{\rm gas}.
    \label{eq:yx}
\end{equation}
Finally, the X-ray luminosity is calculated using:
\begin{equation}
    L_{\rm X} = \sum_im_{{\rm gas},i}\rho_{{\rm gas},i}T_i^{1/2},
    \label{eq:lx}
\end{equation}
where $\rho_{{\rm gas},i}$ is the gas density of gas cell $i$ and the summation is performed over the same gas cells as for the $\bar{T}_{\rm gas}$ calculation.

\begin{figure*}
\centering
\includegraphics[width=1.0\textwidth]{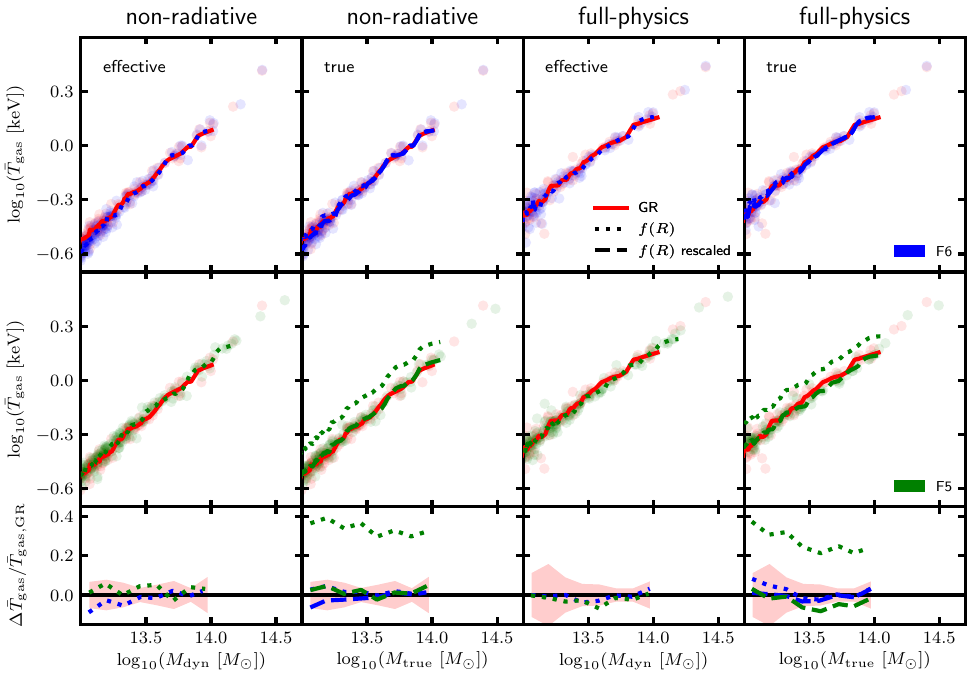}
\caption{[{\it Colour Online}] Gas temperature plotted as a function of the mass for FOF haloes from the non-radiative and full-physics SHYBONE simulations (see Sec.~\ref{sec:simulations}). The curves correspond to the median mass-weighted temperature and the mean logarithm of the mass computed within a moving window of fixed size equal to 10 haloes. Data has been included for GR (\textit{red solid lines}) together with the F6 (\textit{blue lines}) and F5 (\textit{green lines}) $f(R)$ gravity models. Rescalings to the $f(R)$ gravity temperature have been carried out as described in Secs.~\ref{sec:eff_approach} and \ref{sec:true_approach}. For the `true' density approach, the rescaled data (\textit{dashed lines}) is shown along with the unaltered data (\textit{dotted lines}). For this data, the mass corresponds to the total true mass within the radius $R_{500}^{\rm true}$, and the temperature has also been computed within this radius. For the `effective' density approach, no rescaling is necessary, the mass corresponds to the total dynamical mass within $R_{500}^{\rm eff}$, and the temperature has also been computed within this radius. Data points are displayed, with each point corresponding to a GR halo (\textit{red points}) or to a halo in F6 (\textit{blue points}) or F5 (\textit{green points}), including the rescaling for the `true' density data. \textit{Bottom row}: the smoothed relative difference between the $f(R)$ gravity and GR curves in the above plots; the red shaded regions indicate the size of the halo scatter in GR.}
\label{fig:T_gas}
\end{figure*}

\begin{figure*}
\centering
\includegraphics[width=1.0\textwidth]{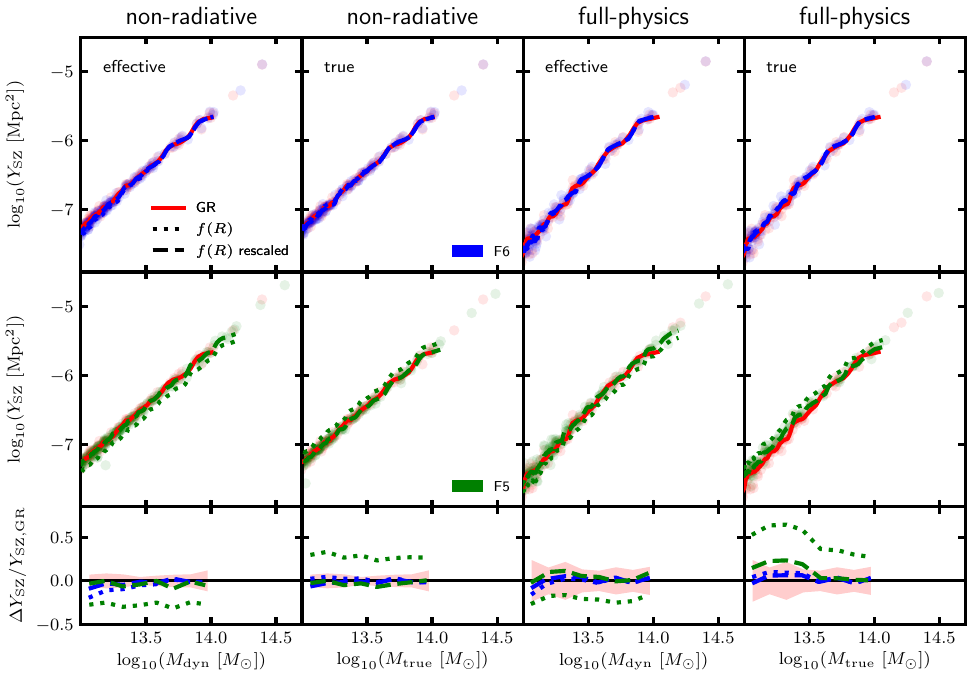}
\caption{[{\it Colour Online}] Compton $Y$-parameter of the SZ effect, plotted as a function of the mass for FOF haloes from the non-radiative and full-physics SHYBONE simulations (see Sec.~\ref{sec:simulations}). The curves correspond to the median $Y_{\rm SZ}$ versus the mean logarithm of the mass computed within a moving window of fixed size equal to 10 haloes. Data has been included for GR (\textit{red solid lines}) together with the F6 (\textit{blue lines}) and F5 (\textit{green lines}) $f(R)$ gravity models. Rescalings to $Y_{\rm SZ}$ have been carried out as described in Sec.~\ref{sec:background:scaling_relations}, and both the rescaled (\textit{dashed lines}) and unaltered (\textit{dotted lines}) results are shown. For the `effective' density approach (see Sec.~\ref{sec:eff_approach}), the mass corresponds to the total dynamical mass within $R_{500}^{\rm eff}$, and $Y_{\rm SZ}$ has also been computed within this radius. For the `true' density approach (see Sec.~\ref{sec:true_approach}), the mass corresponds to the total true mass within the radius $R_{500}^{\rm true}$, and $Y_{\rm SZ}$ has also been computed within this radius. Data points are displayed, with each point corresponding to a GR halo (\textit{red points}), or to a halo in F6 (\textit{blue points}) or F5 (\textit{green points}) with the relevant rescaling applied to $Y_{\rm SZ}$. \textit{Bottom row}: the smoothed relative difference between the $f(R)$ gravity and GR curves in the above plots; the red shaded regions indicate the size of the halo scatter in GR.}
\label{fig:Ysz_scaling_relation}
\end{figure*}

\begin{figure*}
\centering
\includegraphics[width=1.0\textwidth]{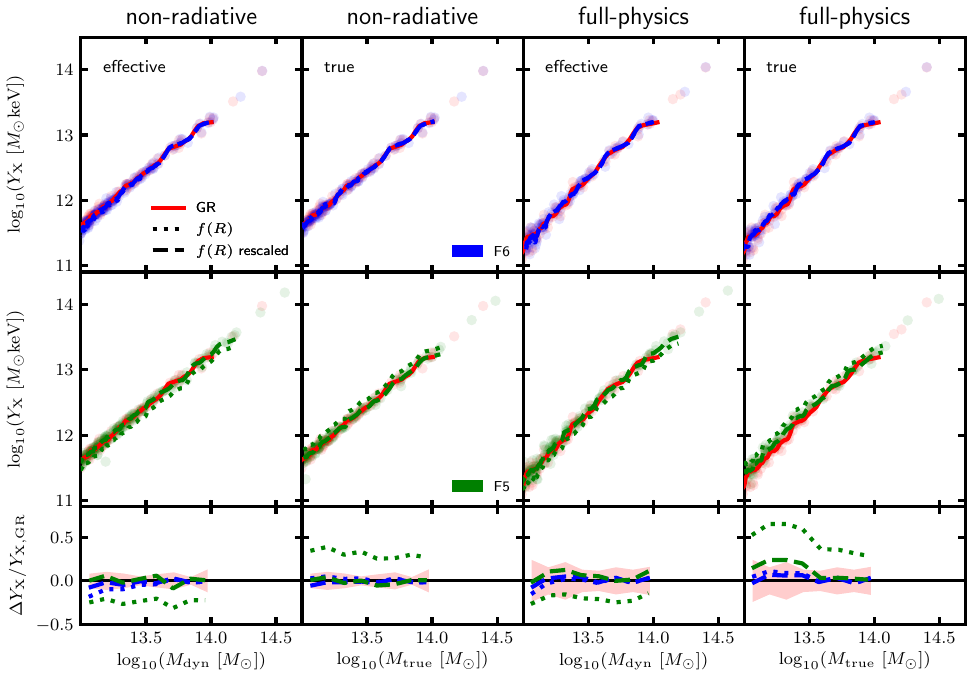}
\caption{[{\it Colour Online}] X-ray analogue of the Compton $Y$-parameter plotted as a function of the mass for FOF haloes from the non-radiative and full-physics SHYBONE simulations (see Sec.~\ref{sec:simulations}). The curves correspond to the median $Y_{\rm X}$ versus the mean logarithm of the mass computed within a moving window of fixed size equal to 10 haloes. Data has been included for GR (\textit{red solid lines}) together with the F6 (\textit{blue lines}) and F5 (\textit{green lines}) $f(R)$ gravity models. Rescalings to $Y_{\rm X}$ have been carried out as described in Sec.~\ref{sec:background:scaling_relations}, and both the rescaled (\textit{dashed lines}) and unaltered (\textit{dotted lines}) results are shown. For the `effective' density approach (see Sec.~\ref{sec:eff_approach}), the mass corresponds to the total dynamical mass within $R_{500}^{\rm eff}$, and $Y_{\rm X}$ has also been computed within this radius. For the `true' density approach (see Sec.~\ref{sec:true_approach}), the mass corresponds to the total true mass within the radius $R_{500}^{\rm true}$, and $Y_{\rm X}$ has also been computed within this radius. Data points are displayed, with each point corresponding to a GR halo (\textit{red points}), or to a halo in F6 (\textit{blue points}) or F5 (\textit{green points}) with the relevant rescaling applied to $Y_{\rm X}$. \textit{Bottom row}: the smoothed relative difference between the $f(R)$ gravity and GR curves in the above plots; the red shaded regions indicate the size of the halo scatter in GR.}
\label{fig:Yx_scaling_relation}
\end{figure*}

\begin{figure*}
\centering
\includegraphics[width=1.0\textwidth]{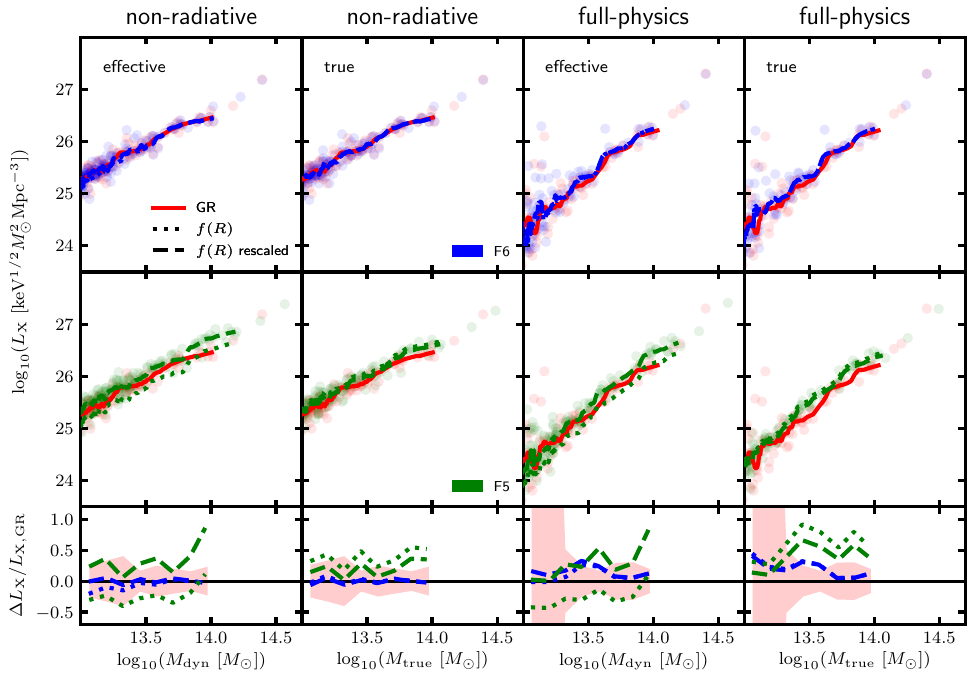}
\caption{[{\it Colour Online}] X-ray luminosity plotted as a function of the mass for FOF haloes from the non-radiative and full-physics SHYBONE simulations (see Sec.~\ref{sec:simulations}). The curves correspond to the median luminosity versus the mean logarithm of the mass computed within a moving window of fixed size equal to 10 haloes. Data has been included for GR (\textit{red solid lines}) together with the F6 (\textit{blue lines}) and F5 (\textit{green lines}) $f(R)$ gravity models. Rescalings to the luminosity have been carried out as described in Sec.~\ref{sec:background:scaling_relations}, and both the rescaled (\textit{dashed lines}) and unaltered (\textit{dotted lines}) results are shown. For the `effective' density approach (see Sec.~\ref{sec:eff_approach}), the mass corresponds to the total dynamical mass within $R_{500}^{\rm eff}$, and the luminosity has also been computed within this radius. For the `true' density approach (see Sec.~\ref{sec:true_approach}), the mass corresponds to the total true mass within the radius $R_{500}^{\rm true}$, and the luminosity has also been computed within this radius. Data points are displayed, with each point corresponding to a GR halo (\textit{red points}), or to a halo in F6 (\textit{blue points}) or F5 (\textit{green points}) with the relevant rescaling applied to the luminosity. \textit{Bottom row}: the smoothed relative difference between the $f(R)$ gravity and GR curves in the above plots; the red shaded regions indicate the size of the halo scatter in GR.}
\label{fig:Lx_scaling_relation}
\end{figure*}

\section{Results}
\label{sec:results}

In Sec.~\ref{sec:results:scaling_relations}, we discuss our results for the cluster scaling relations in HS $f(R)$ gravity. Then, in Sec.~\ref{sec:results:mass_ratio}, we test the validity of our analytical tanh formula for the dynamical mass enhancement, given by Eq.~(\ref{eq:enhancement}), in the presence of full physics. There we will also present an example in which we map between the GR and $f(R)$ mass scaling relations based on this approximate fitting formula, rather than the actual values of $M^{f(R)}_{\rm dyn}/M^{f(R)}_{\rm true}$ from the simulations. Finally, in Sec.~\ref{sec:results:yx-t_relation}, we propose a new test of gravity using the $Y_{\rm X}$-$\bar{T}_{\rm gas}$ relation, which does not require inferences of the cluster mass.

\subsection{Scaling relations}
\label{sec:results:scaling_relations}

Using our simulation data, we have tested the scaling relation mappings described in Sec.~\ref{sec:background:scaling_relations}. Due to the small box size, $62h^{-1}{\rm Mpc}$ (comoving), of our simulations, we can only examine haloes with mass $M_{500}\lesssim10^{14.5}M_{\odot}$. We show all objects with $M_{500}\geq10^{13}M_{\odot}$ (groups and clusters) in Figs.~\ref{fig:T_gas}--\ref{fig:Lx_scaling_relation}, which typically includes $\sim100$ haloes for a given model. Note that it is difficult to rigorously test our scaling relation mappings for the cluster regime ($M_{500}\gtrsim10^{14}M_{\odot}$), where there are only 5-10 haloes in the present simulations. However, for F5, groups are typically unscreened and low-mass clusters are partially screened, while for F6 low-mass groups are partially screened and higher-mass objects are completely screened; haloes with $M_{500}\gtrsim10^{14.5}M_{\odot}$ will be mostly screened for F5 and entirely screened for F6 \citep[see, e.g., Fig.~\ref{fig:mdyn_mtrue} below and Fig.~5 of][]{Mitchell:2018qrg}. Therefore, while we do not have a significant number of such large cluster-sized objects, we do expect the scaling relations calibrated for GR to be valid for them.

In addition to showing data points for individual haloes, all plots include curves showing a moving average. This is computed using a moving window of fixed size equal to 10 haloes, for which the mean logarithm of the mass and the median proxy are displayed. We note that the highest-mass haloes have been included in the moving average, even though the highest mean mass is only $\sim10^{14}M_{\odot}$. We also show sub-plots with the smoothed relative difference between the $f(R)$ and GR curves, as well as the halo scatter in GR. The latter is computed by fitting a linear model to the GR data and computing the root-mean-square residuals within mass bins.

For the panels labelled `effective' in Figs.~\ref{fig:T_gas}--\ref{fig:Lx_scaling_relation}, all measurements of the mass and observables have been taken within $R_{500}^{\rm eff}$ (see Sec.~\ref{sec:eff_approach}), and the relations are plotted against the dynamical mass. This allows the effective density mappings given by Eqs.~(\ref{eq:temp_equiv_eff}) and (\ref{eq:ysz_mapping})-(\ref{eq:lx_mapping}), originally proposed by \citet{He:2015mva}, to be tested. On the other hand, an outer radius $R_{500}^{\rm true}$ (see  Sec.~\ref{sec:true_approach}) is imposed for all measurements for the data displayed in the panels labelled `true'. These are plotted against the true mass, and can be used to test the true density mappings given by Eqs.~(\ref{eq:temp_equiv_true}) and (\ref{eq:ysz_mapping_true})-(\ref{eq:lx_mapping_true}).

\subsubsection{Temperature scaling relations}
\label{sec:results:tgas}

The results for the gas temperature scaling relations are shown in Fig.~\ref{fig:T_gas}. The non-radiative data is displayed in the left two columns and the full-physics data is shown in the right two columns. For all models and hydrodynamical schemes, the data follows a power-law behaviour, as expected from the self-similar model. The correlation is particularly tight for the non-radiative data, with an overall scatter of 7\%. The non-radiative runs contain gas and dark matter particles, but do not feature baryonic processes (apart from basic hydrodynamics) such as radiative cooling, stellar and black hole feedback and star formation. It is therefore expected that the thermodynamical properties can be largely determined from the gravitational potential, which is observed in the results. On the other hand, there is $\sim10\%$ overall scatter in the full-physics data, and the gas temperature is typically higher. This can be explained by the inclusion of feedback mechanisms which act as an additional source of heating of the surrounding gas and cause some departures from self-similarity. These mechanisms have a stronger effect on lower-mass haloes, resulting in a particularly high ($10$-$20\%$) scatter for these objects.

For the effective density approach, Eq.~(\ref{eq:temp_equiv_eff}) is expected to hold: the temperature is predicted to be equal for haloes in GR and $f(R)$ gravity with the same dynamical mass. In Fig.~\ref{fig:T_gas}, the non-radiative and full-physics results from our effective catalogue are shown in the first and third columns from the left, respectively. For both F6 and F5, there is excellent agreement with the GR data, with typical agreement $\lesssim5\%$. This agreement for the non-radiative data backs up the findings from \citet{He:2015mva}, while the full-physics results do not show clear evidence for a departure from Eq.~(\ref{eq:temp_equiv_eff}) caused by feedback processes and cooling. These results agree with the self-similar model predictions: two haloes in $f(R)$ gravity and GR which have the same dynamical mass $M_{\rm dyn}$ (and therefore the same radius $R_{500}^{\rm eff}$) also have the same gravitational potential, $\phi = GM_{\rm dyn}/R_{500}^{\rm eff}$.

In order to test the new mappings predicted by the true density approach, the temperature of each halo in $f(R)$ gravity has been divided by the mass ratio $M_{\rm dyn}^{f(R)}/M_{\rm true}^{f(R)}$. In Fig.~\ref{fig:T_gas}, both the data for individual haloes and corresponding moving averages are shown with this rescaling applied (dashed lines), along with the moving averages for the unaltered data (dotted lines). It is expected, from Eq.~(\ref{eq:temp_equiv_true}), that the data with the rescaling should agree with GR. For the non-radiative results in the second column, this indeed appears to be the case, with an excellent agreement that is generally within just a few percent. For the full-physics data there is still reasonable $\lesssim10\%$ agreement, but the F5 temperature appears to be lower than the GR temperature for $\log_{10}\left[M_{\rm true}\left(<R_{500}^{\rm true}\right)/M_{\odot}\right]\gtrsim13.5$.

This small deviation is consistent with the full-physics temperature profiles shown in Fig.~\ref{fig:profiles}. The plots in the bottom right of that figure show the temperature profiles with the $M_{\rm dyn}^{f(R)}/M_{\rm true}^{f(R)}$ rescaling applied. The profiles are shown for the halo mass bins $10^{13.7}M_{\odot}<M_{\rm true}\left(<R_{500}^{\rm true}\right)<10^{14.0}M_{\odot}$ and $10^{13.0}M_{\odot}<M_{\rm true}\left(<R_{500}^{\rm true}\right)<10^{13.3}M_{\odot}$. 
For the high-mass full-physics profiles, the rescaled F5 profile is clearly lower than the profile in GR across most of the radial range. This can explain the lower rescaled F5 temperature observed at the high-mass end of the full-physics data. On the other hand, for the lower-mass bin with full-physics and for the non-radiative data the agreement between the rescaled $f(R)$ gravity and GR temperature profiles is very good, particularly at the outer radii which have greater overall contribution to the mass-weighted temperature. Similar agreement is shown between the $f(R)$ gravity and GR profiles for the plots in the bottom-left of Fig.~\ref{fig:profiles}. Again, for the high-mass full-physics data there is some deviation between F5 and GR, particularly in the outermost regions. But this is not as noticeable as for the profiles with the true density rescalings.

The small difference between F5 and GR is likely to be caused by a difference in the levels of feedback -- which itself is determined by the interrelations between modified gravity (including screening or the lack of it) and baryonic physics -- in these higher-mass haloes for the two models. Encouragingly, this appears to have only a small effect on the effective density data, where there is good agreement between F5 and GR for high-mass groups. However, the effect is greater for the true density data. To understand the implications that this could have on tests of gravity using the cluster regime, we will need simulations with a larger box size. This is planned as a future work, where we will use a re-calibrated full-physics model to probe masses up to $M_{500}\sim10^{15}M_{\odot}$. Although the current simulations are unable to rigorously probe cluster-sized objects ($M_{500}\gtrsim10^{14}M_{\odot}$), it does appear that the relative difference curve approaches zero at $M_{500}\sim10^{14}M_{\odot}$. The planned large-box simulation will allow us to study the interplay between the (partially screened) fifth force and baryonic feedbacks in greater details.

\subsubsection{$Y_{\rm SZ}$ and $Y_{\rm X}$ scaling relations}
\label{sec:results:y_params}

Our results for the $Y_{\rm SZ}$ and $Y_{\rm X}$ scaling relations are shown in Figs.~\ref{fig:Ysz_scaling_relation} and \ref{fig:Yx_scaling_relation}. The $Y_{\rm SZ}$ and $Y_{\rm X}$ parameters are, by definition, tightly correlated. Their results therefore follow similar patterns, and both show very tight correlations with the halo mass, with a scatter of $\sim8\%$ and $\sim19\%$ for the non-radiative and full-physics data, respectively. There are also no clear outliers in the full-physics data, unlike for the temperature and the X-ray luminosity data (see below). This is because of the competing effects of feedback processes on the gas density and gas temperature \citep{Fabjan:2011}. Comparing the non-radiative and full-physics profiles in Fig.~\ref{fig:profiles}, it can be seen that the additional processes in the full-physics runs cause haloes to have a lower gas density, particularly at the inner regions, and a higher gas temperature. This is caused by stellar and black hole feedbacks, which generate high-energy winds that heat up the surrounding gas and blow it out from the central regions. Such competing effects are approximately
cancelled out in the product of the gas density with the gas temperature, as in Eqs.~(\ref{eq:ysz_obs}) and (\ref{eq:yx_obs}). 

In order to test the mappings predicted by the effective density approach, given by Eqs.~(\ref{eq:ysz_mapping}) and (\ref{eq:yx_mapping}), the $Y_{\rm SZ}$ and $Y_{\rm X}$ values measured for $f(R)$ gravity have been multiplied by the mass ratio $M_{\rm dyn}^{f(R)}/M_{\rm true}^{f(R)}$. For the non-radiative plots in Figs.~\ref{fig:Ysz_scaling_relation} and \ref{fig:Yx_scaling_relation}, there is excellent agreement between this rescaled data and GR. There is also a strong agreement for the higher-mass full-physics data. For the mass range $13.2<\log_{10}(M_{\rm dyn}(<R_{500}^{\rm eff})M_{\odot}^{-1})<13.5$, however, there is some disparity of $\lesssim20\%$ between the rescaled F5 data and GR.

A similar level of accuracy is observed for the mappings given by Eqs.~(\ref{eq:ysz_mapping_true}) and (\ref{eq:yx_mapping_true}), which are predicted by the true density approach. To test these, $Y_{\rm SZ}$ and $Y_{\rm X}$ are divided by $M_{\rm dyn}^{f(R)}/M_{\rm true}^{f(R)}$ to generate rescaled data for F6 and F5. Again, this data shows excellent agreement, within a few percent, with GR for the non-radiative simulations and the high-mass end of the full-physics data. But for the lower-mass full-physics data there is a significant disagreement between F5 and GR of up to $\sim30\%$, which is higher than for the effective density rescalings. 

The disparities found in the low-mass full-physics data can be explained using Fig.~\ref{fig:profiles}. Looking at the full-physics profiles for the true density mass bins, it is observed that for a large portion of the inner halo regions the gas density is higher in F5 than in GR. These profiles converge at $r\approx10^{2.5}{\rm kpc}$ for both mass bins. The high-mass haloes have higher overall radius $R_{500}^{\rm true}$, which means that the profiles are converged for a large portion of the outer radii. This means the disparities at lower radii have a negligible overall contribution to the integrals for $Y_{\rm SZ}$ and $Y_{\rm X}$. But for the lower-mass haloes, the profiles are converged only for a small portion of the overall radius range, causing $Y_{\rm SZ}$ and $Y_{\rm X}$ to be greater in F5 than in GR. A similar reasoning can be used to explain the disparities for the results with the effective density rescaling, although the difference in agreement at the outer regions for each mass bin is not quite as substantial here, which explains why the full-physics data for the effective density approach shows less overall deviation between F5 and GR in Figs.~\ref{fig:Ysz_scaling_relation} and \ref{fig:Yx_scaling_relation}.

The difference between the full-physics $f(R)$ and GR scaling relations at low mass is likely to be explained by baryonic processes such as feedback which are absent in the non-radiative simulations. However, in studies of clusters, these lower-mass groups are of less interest. The strong agreement at the higher masses is therefore very encouraging for our framework to constrain $f(R)$ gravity using the high-mass end of the halo mass function.

\subsubsection{X-ray luminosity scaling relations}
\label{sec:results:lx}

The results for the X-ray luminosity scaling relations are shown in Fig.~\ref{fig:Lx_scaling_relation}. Compared with the temperature, $Y_{\rm SZ}$ and $Y_{\rm X}$ data, the X-ray luminosity is much more scattered, with particularly large dispersion in lower-mass haloes and $\sim25\%$ scatter at higher masses. One explanation for this is that the X-ray luminosity, defined in Eq.~(\ref{eq:lx_obs}), depends on the gas density to power two. This means that the inner regions of the group, which have a higher gas density than the outer regions, have a greater overall contribution to the $L_{\rm X}$ integral than for the other observables discussed in this work. The inner halo regions are expected to be more impacted by unpredictable dynamical processes during cluster formation, including halo mergers. In particular, they are more prone to gas heating and blowing-out of gas caused by feedback mechanisms. While the competing effects of these processes on the gas density and gas temperature roughly cancel for the $Y_{\rm SZ}$ and $Y_{\rm X}$ observables, this is not the case for $L_{\rm X}$, which depends on the gas density to power two and the gas temperature to power half. This results in a number of significant outliers, as can be seen in the full-physics data of Fig.~\ref{fig:Lx_scaling_relation}. 

For the mapping defined using the effective density field, given by Eq.~(\ref{eq:lx_mapping}), it has been expected that the GR X-ray luminosity should be equal to the $f(R)$ gravity value multiplied by the factor $\left(M_{\rm dyn}^{f(R)}/M_{\rm true}^{f(R)}\right)^2$. From Fig.~\ref{fig:Lx_scaling_relation}, the rescaled data in F5 appears to be higher than in GR by $\sim30\%$ on average for both the non-radiative and the full-physics simulations. A similar level of deviation is also observed for the true density results, where Eq.~(\ref{eq:lx_mapping_true}) predicts that the GR and $f(R)$ gravity X-ray luminosity should be equal after the values in $f(R)$ gravity are divided by the factor $\left(M_{\rm dyn}^{f(R)}/M_{\rm true}^{f(R)}\right)^{1/2}$. Again, the rescaled F5 X-ray luminosity is significantly greater than in GR on average. 

As for the $Y_{\rm SZ}$ and $Y_{\rm X}$ mappings, the disparity found here can be explained by looking at the gas density profiles in Fig.~\ref{fig:profiles}. For both the non-radiative and full-physics data, the gas density in the inner halo regions is greater for F5 (with appropriate rescaling applied) than for GR. Because the inner regions have a greater contribution to the X-ray luminosity than for other proxies, as described above, this causes these differences in the inner regions to become significant overall, even for the non-radiative data for which the F5 and GR profiles are converged above a lower radius. This results in the general offset for the full range of masses as shown in Fig.~\ref{fig:Lx_scaling_relation}. As described above, the X-ray luminosity is also more strongly influenced by feedback processes, which can further increase the offset between F5 and GR if the feedback behaves differently in these two models.

Our observation that the mappings have a poorer performance for the X-ray luminosity than for the other proxies is consistent with \citet{He:2015mva}, who observed a disparity of $\sim13\%$. Unless further corrections are applied to account for the unpredictable effects of feedback in the mappings, therefore, the X-ray luminosity is unlikely to be a reliable proxy for mass determination in accurate cluster tests of $f(R)$ gravity.

\subsubsection{Further comments}

As described in Sec.~\ref{sec:methods:groups}, we have excluded the core region of $r<0.15R_{500}$ when calculating the thermal properties of our simulated FOF groups. We have also experimented excluding core regions of size $r<0.1R_{500}$ and $r<0.2R_{500}$: for the scaling relations $Y_{\rm SZ}$-$M$, $Y_{\rm X}$-$M$ and $\bar{T}_{\rm gas}$-$M$, the relative differences between $f(R)$ gravity and GR are barely affected; on the other hand, the effect is larger for the $L_{\rm X}$-$M$ scaling relation, because $L_{\rm X}$ is more sensitive to the inner halo regions than the other proxies. However, the $L_{\rm X}$-$M$ relation is not ideal for reliable tests of gravity anyway, as noted above, and is mainly included in this work for completeness and for comparison with the other relations. Even if the entire core is included in the calculations, we have found that the effect on the $Y_{\rm SZ}$-$M$ and $Y_{\rm X}$-$M$ results is still very small, providing further confirmation that these relations can be used for reliable tests of gravity.

The scatter of the full-physics GR scaling relations shown in Figs.~\ref{fig:T_gas}--\ref{fig:Lx_scaling_relation} is typically higher than the scatter quoted in recent studies that have also used simulations which include star formation, cooling and stellar and black hole feedback. For example, we observe a root-mean-square dispersion of $\sim19\%$ for the $Y_{\rm SZ}$-$M$ and $Y_{\rm X}$-$M$ relations, while \citet{Brun:2016jtk} and \citet{Truong:2016egq} reported $\sim10\%$ and $\sim15\%$, respectively. This is likely to be caused by our restricted halo population (modified gravity simulations are much more computationally expensive than their standard gravity counterparts which limits the affordable box-size and resolution), which contains a large number of low-mass groups that are more susceptible to feedback. Our results suggest that the $\bar{T}_{\rm gas}$-$M$ relation has the lowest scatter and the $L_{\rm X}$-$M$ relation has the highest scatter, and this is consistent with the above works.

\begin{figure*}
\centering
\includegraphics[width=1.0\textwidth]{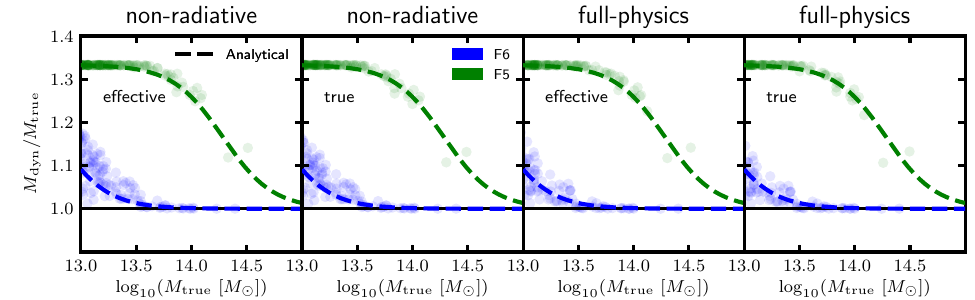}
\caption{[{\it Colour Online}] Ratio of the dynamical mass to the true mass of haloes plotted as a function of the true mass. The data points correspond to FOF haloes from the non-radiative and full-physics SHYBONE simulations (see Sec.~\ref{sec:simulations} for details). Data is included for F6 (\textit{blue}) and F5 (\textit{green}), which are the HS $f(R)$ gravity models with $n=1$ and present-day background scalar field $|f_{R0}|=10^{-6}$ and $|f_{R0}|=10^{-5}$, respectively. For the data labelled `effective', the dynamical and true mass have been measured within the radius $R_{500}^{\rm eff}$ (defined in Sec.~\ref{sec:eff_approach}), while the radius $R_{500}^{\rm true}$ (defined in Sec.~\ref{sec:true_approach}) has been used for the data labelled `true'. Analytical predictions for the mass enhancement have been computed using Eq.~(\ref{eq:enhancement}) and are shown (\textit{dashed curves}) for each model.}
\label{fig:mdyn_mtrue}
\end{figure*}

\begin{figure*}
\centering
\includegraphics[width=1.0\textwidth]{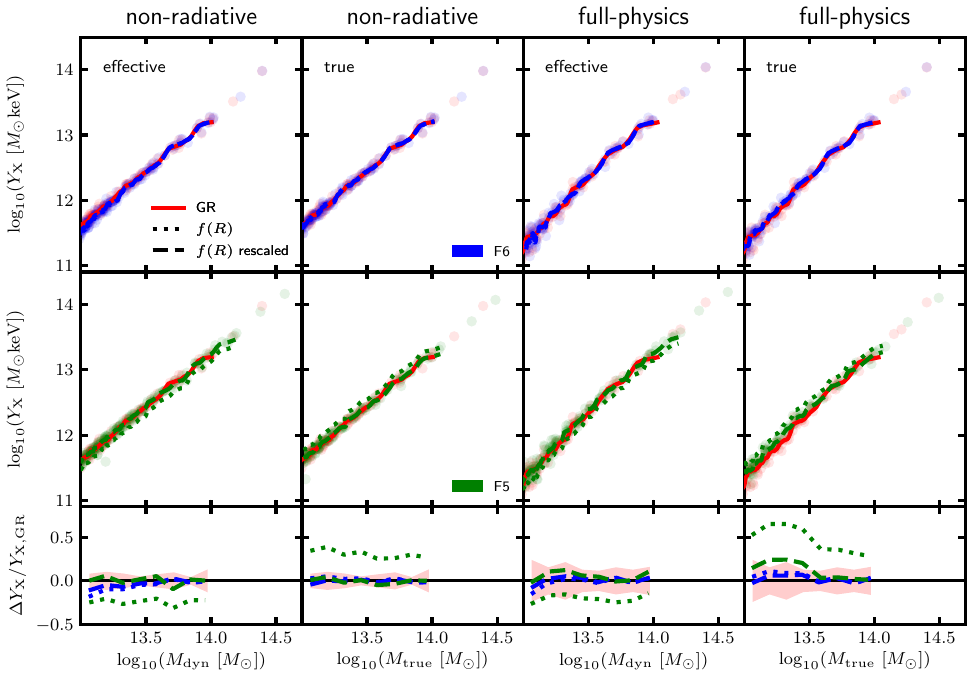}
\caption{[{\it Colour Online}] X-ray analogue of the Compton $Y$-parameter plotted as a function of the mass for FOF haloes from the non-radiative and full-physics SHYBONE simulations (see Sec.~\ref{sec:simulations}). The layout and format of this figure are identical to those of Fig.~\ref{fig:Yx_scaling_relation}. The results shown here are also mostly the same as for Fig.~\ref{fig:Yx_scaling_relation}, however the rescalings of $Y_{\rm X}$ in the $f(R)$ gravity data have been generated using an analytical tanh formula, given by Eq.~(\ref{eq:enhancement}).}
\label{fig:Yx_scaling_relation_TANH}
\end{figure*}

\subsection{Halo mass ratio calculation}
\label{sec:results:mass_ratio}

For the results discussed in Sec.~\ref{sec:results:scaling_relations}, the rescalings to the observables in $f(R)$ gravity have been computed using direct measurements of the true mass and the dynamical mass from the simulations. However, for studies of clusters using real observations, measurements of both the true mass and dynamical mass are unlikely to be available. In this case, our analytical model for the ratio of the dynamical mass to the true mass, given by Eq.~(\ref{eq:enhancement}), can be used. This formula was calibrated by \citet{Mitchell:2018qrg} using a suite of dark-matter-only simulations and tested for wide and continuous ranges of scalar field values, redshifts and halo masses.

To check how this model performs for data that includes full physics, we have plotted this on top of actual measurements of the dynamical mass enhancement for the FOF groups in the SHYBONE simulations. This is shown in Fig.~\ref{fig:mdyn_mtrue}. Data for both the effective density and true density catalogues have been included, for which the dynamical and true halo masses have been measured within $R_{500}^{\rm eff}$ and $R_{500}^{\rm true}$, respectively. We have made use of all available data with $M_{\rm true}>10^{13}M_{\odot}$, including haloes with $M_{\rm true}\sim10^{14.5}M_{\odot}$. These results indicate that there is very good agreement between the analytical predictions and the actual data for both F6 and F5, regardless of the hydrodynamical scheme that is employed. Interestingly, even though Eq.~(\ref{eq:enhancement}) was originally calibrated using measurements of the dynamical and true mass within $R_{500}^{\rm true}$, it still performs very well for data measured within $R_{500}^{\rm eff}$, which is typically a higher radius.

We have also tested the mappings of Eqs.~(\ref{eq:yx_mapping}) and (\ref{eq:yx_mapping_true}) for the $Y_{\rm X}$ parameter, with Eq.~(\ref{eq:enhancement}) used to compute the required rescalings to the $f(R)$ gravity data. This is shown in Fig.~\ref{fig:Yx_scaling_relation_TANH}. From comparing this plot with Fig.~\ref{fig:Yx_scaling_relation}, it can be seen that there is almost no difference in the rescaled data in both figures. This confirms that Eq.~(\ref{eq:enhancement}) can be applied to derive the mappings between GR and $f(R)$ scaling relations for, at least, the mass range $10^{13}M_{\odot}<M_{500}<10^{14}M_{\odot}$. Given the very good agreement up to $10^{14.5}M_{\odot}$ shown in Fig.~\ref{fig:mdyn_mtrue}, it is expected that our formula can be applied in the cluster regime as well. We will test this more rigorously in a future study by using full-physics simulations with a larger box size.

\begin{figure*}
\centering
\includegraphics[width=1.0\textwidth]{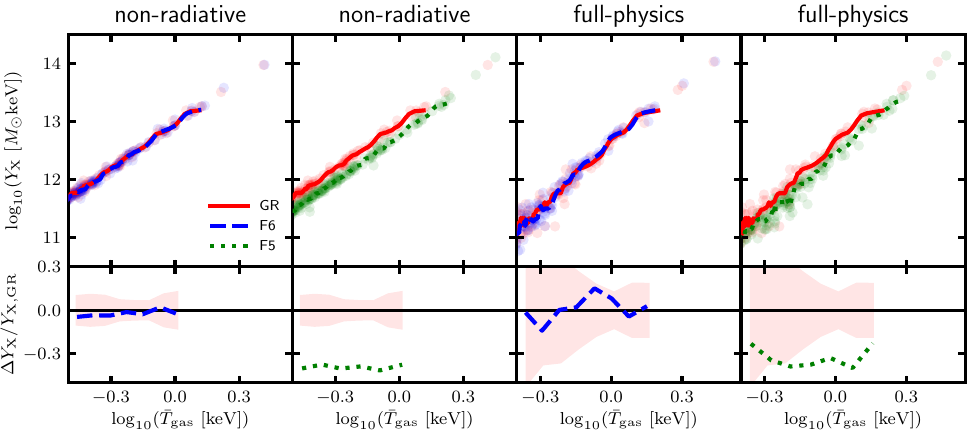}
\caption{[{\it Colour Online}] The X-ray analogue of the Compton $Y$-parameter plotted as a function of the mass-weighted temperature for FOF haloes from the non-radiative and full-physics SHYBONE simulations (see Sec.~\ref{sec:simulations}). The curves correspond to the median $Y_{\rm X}$ versus the mean logarithm of the temperature computed within a moving window of fixed size equal to 10 haloes. Data has been included for GR (\textit{red solid lines}) together with the F6 (\textit{blue lines}) and F5 (\textit{green lines}) $f(R)$ gravity models. $Y_{\rm X}$ and the temperature have been computed within the radius $R_{500}^{\rm true}$ (see Sec.~\ref{sec:true_approach}). Data points are displayed, with each point corresponding to a GR halo (\textit{red points}), or to a halo in F6 (\textit{blue points}) or F5 (\textit{green points}). \textit{Bottom row}: the smoothed relative difference between the $f(R)$ gravity and GR curves in the above plots; the red shaded regions indicate the size of the halo scatter in GR.}
\label{fig:Yx-T_scaling_relation}
\end{figure*}

\subsection{\boldmath $Y_{\rm X}$-temperature scaling relation}
\label{sec:results:yx-t_relation}

So far, we have only considered scaling relations that can be used to infer the cluster mass, which is a vital ingredient for tests of gravity that use the cluster abundance (Fig.~\ref{fig:flow_chart}). However, tests of gravity can also be conducted using the relations themselves. For example, \citet{Hammami:2016npf,DelPopolo:2019oxn} used the temperature-mass scaling relation to probe screened MG models. The cluster mass can be determined using other observations, such as weak lensing, making such scaling relations observable, and our full-physics results (Figs.~\ref{fig:T_gas}--\ref{fig:Yx_scaling_relation}) confirm that the $\bar{T}_{\rm gas}$-$M$ and $Y_{\rm X}$-$M$ (and $Y_{\rm SZ}$-$M$) relations can be used as reliable probes on group and cluster scales, with differences between GR and F5 in the range 20-50\% (when no rescaling is applied).

However, scaling relations which do not involve the mass can also be modelled and used. In Fig.~\ref{fig:Yx-T_scaling_relation}, we show the scaling relation between the $Y_{\rm X}$ parameter and the gas temperature, where both observables have been computed within the radius $R_{500}^{\rm true}$. A significant relative difference of 30-40\% is observed between the GR and F5 models for both the non-radiative and full-physics data, indicating that this relation can offer a powerful test of gravity using group- and cluster-sized objects. A key advantage of using the $Y_{\rm X}$--$\bar{T}_{\rm gas}$ (or $Y_{\rm SZ}$--$\bar{T}_{\rm gas}$) scaling relation is that it does not involve measuring the cluster mass, and hence no need for mass calibrations or synergies with other observations such as weak lensing. It also has a relatively low scatter compared to, for example, the $L_{\rm X}$-$\bar{T}_{\rm gas}$ relation that was considered by \citet{arnold:2014}.

\section{Summary, Discussion and Conclusions}
\label{sec:conclusions}

Accurate determination of the cluster mass is a vital requirement of tests that make use of LSS probes including cluster counts to constrain cosmological models. It is particularly important in tests of chameleon-type MG theories, in which the additional forces present can alter the internal properties of clusters, including the concentration, the dynamical mass and the temperature. This means that scaling relations between the cluster mass and observable proxies that have been determined assuming GR are unlikely to apply in a universe with, for example, chameleon $f(R)$ gravity.

This work is part of a series \citep{Mitchell:2018qrg,Mitchell:2019qke} which aims to create the first complete and robust pipeline for accounting for the effects of $f(R)$ gravity on the internal properties of galaxy clusters. Highlights so far include the calibration of analytical models that can accurately describe the enhancement of the dynamical mass \citep{Mitchell:2018qrg} and concentration \citep{Mitchell:2019qke} of haloes in HS $f(R)$ gravity. These models are accurate for a wide and continuous range of masses, redshifts and values of the background scalar field. In particular, we have found a powerful way to describe the chameleon screening mechanism with just a single parameter: $|f_R|/(1+z)$, and in a future work will use this to calibrate a general model for the halo mass function in $f(R)$ gravity.

In this work, we have made use of the first full-physics simulations that have been run for both GR and $f(R)$ gravity (along with non-radiative counterparts), to study the effects of the fifth force of $f(R)$ gravity on the scaling relations between the cluster mass and four observable proxies: the gas temperature (Fig.~\ref{fig:T_gas}), the $Y_{\rm SZ}$ and $Y_{\rm X}$ parameters (Figs.~\ref{fig:Ysz_scaling_relation} and \ref{fig:Yx_scaling_relation}) and the X-ray luminosity (Fig.~\ref{fig:Lx_scaling_relation}). To understand these effects in greater detail, we have also examined the effects of both $f(R)$ gravity and full-physics on the gas density and temperature profiles (see Fig.~\ref{fig:profiles}). In doing so, we have been able to test two 
methods for mapping between scaling relations in $f(R)$ gravity and GR.

The first method was proposed by \citet{He:2015mva}. This proposes a set of mappings, given by Eqs.~(\ref{eq:temp_equiv_eff}) and (\ref{eq:ysz_mapping})-(\ref{eq:lx_mapping}), that can be applied to haloes whose mass and radius are measured using the effective density field (see Sec.~\ref{sec:eff_approach}). A second, new, approach is proposed in Sec.~\ref{sec:true_approach}, and predicts another set of mappings, given by Eqs.~(\ref{eq:temp_equiv_true}) and (\ref{eq:ysz_mapping_true})-(\ref{eq:lx_mapping_true}), that can be applied to haloes whose mass and radius are measured using the true density field. Both sets of mappings involve simple rescalings that depend only on the ratio of the dynamical mass to the true mass in $f(R)$ gravity. As shown by Figs.~\ref{fig:mdyn_mtrue} and \ref{fig:Yx_scaling_relation_TANH}, even with the inclusion of full-physics processes this ratio can be computed with high accuracy using our analytical tanh formula, which is given by Eq.~(\ref{eq:enhancement}).

For the mass-weighted gas temperature and the $Y_{\rm SZ}$ and $Y_{\rm X}$ observables, we found that the F6 and F5 scaling relations, with appropriate rescaling applied (using either method discussed above), match the GR relations to within a few percent for the full mass-range tested for the non-radiative simulations. With the inclusion of full-physics effects such as feedbacks, star formation and cooling, the rescaled $Y_{\rm SZ}$ and $Y_{\rm X}$ scaling relations continue to show excellent agreement with GR for mass $M_{500}\gtrsim10^{13.5}M_{\odot}$, which includes group- and cluster-sized objects. These proxies also show relatively low scatter as a function of the cluster mass, compared with other observables. $Y_{\rm SZ}$ and $Y_{\rm X}$ are therefore likely to be suitable for accurate determination of the cluster mass in tests of $f(R)$ gravity. The mappings for the gas temperature show a very high accuracy for lower-mass objects, but show a small $\lesssim5\%$ offset between F5 and GR for higher-mass objects.

The mappings do not work as well for the X-ray luminosity $L_{\rm X}$, for which the F5 relations after rescaling are typically enhanced by $\sim30\%$ compared with GR. This is caused by the unique dependency of $L_{\rm X}$ on the gas density to power two, and the gas temperature to power half, which means that the inner halo regions have a greater contribution than for the other proxies and the competing effects of feedback on the temperature and gas density profiles are less likely to cancel out. This issue, in addition to the fact that $L_{\rm X}$ has a highly scattered correlation with the cluster mass, means that this proxy is unlikely to be suitable for cluster mass determination in tests of $f(R)$ gravity.

We also considered the $Y_{\rm X}$-$\bar{T}_{\rm gas}$ scaling relation (Fig.~\ref{fig:Yx-T_scaling_relation}), and found that this is suppressed by $30$-$40\%$ in the F5 model relative to GR. This offers a potential new and useful test of gravity with group- and cluster-sized objects which avoids the systematic uncertainties incurred from mass calibration.

We note that the box size $62 h^{-1}{\rm Mpc}$ of the simulations used in this work is more suited to studying galaxy-sized objects than group- or cluster-sized objects. Indeed, there are only $\sim100$ objects with $M_{500}>10^{13}M_{\odot}$ and $\sim5$-$10$ objects with $M_{500}>10^{14}M_{\odot}$ in the simulations. This makes it impossible to test the mappings discussed in this work for the most massive galaxy clusters to be observed. We are therefore currently preparing to run larger simulations, which employ a re-calibrated full-physics model, that can be used to reliably probe halo masses up to $M_{500}\sim10^{15}M_{\odot}$. We will leave the analysis of these simulations to a future work. 

Our results also provide insights into the viability of extending cluster tests of gravity to the group-mass regime. An advantage of using lower-mass objects is that these objects can be unscreened (or partially screened) even for weaker $f(R)$ models, offering the potential for tighter constraints using data from ongoing and upcoming SZ and X-ray surveys \citep[e.g.,][]{erosita,Planck_SZ_cluster} which are now entering this regime. On the other hand, as we have seen above, the scatter induced by feedback mechanisms becomes more significant in group-sized haloes, which means that additional work need to be conducted to characterise this effect and to understand its impact on model tests.

Finally, we note that our parameter $p_2$, which is used to compute the ratio of the dynamical mass to the true mass, depends only on the quantity $|f_R|/(1+z)$, and not on the model parameters $n$ and $f_{R0}$ of HS $f(R)$ gravity. This dependence was originally derived by using the thin-shell model \citep{Mitchell:2018qrg}, which does not depend on the details of the $f(R)$ model. We therefore expect our scaling relation mappings to perform similarly for any combination of the HS $f(R)$ parameters, and potentially other chameleon-type or thin-shell-screened models. However, due to the high computational cost of running full-physics simulations of $f(R)$ gravity and other models, we do not seek to confirm this conjecture in this work. We are also extending our framework to other MG theories, starting with the normal-branch Dvali-Gabadadze-Porrati model \citep{DVALI2000208} using the simulations described in \citet{Hernandez-Aguayo:2020kgq}. We will present all these results in future works.

\section*{Acknowledgements}

We thank Jianhua He for helpful discussions during this project. MAM is supported by a PhD Studentship with the Durham Centre for Doctoral Training in Data Intensive Science, funded by the UK Science and Technology Facilities Council (STFC, ST/P006744/1) and Durham University. CA and BL are supported by the European Research Council via grant ERC-StG-716532-PUNCA. BL is additionally supported by STFC Consolidated Grants ST/T000244/1 and ST/P000541/1. This work used the DiRAC@Durham facility managed by the Institute for Computational Cosmology on behalf of the STFC DiRAC HPC Facility (\url{www.dirac.ac.uk}). The equipment was funded by BEIS capital funding via STFC capital grants ST/K00042X/1, ST/P002293/1, ST/R002371/1 and ST/S002502/1, Durham University and STFC operations grant ST/R000832/1. DiRAC is part of the National e-Infrastructure.

\section*{Data availability}

The simulation data and results of this paper may be available upon request.




\bibliographystyle{mnras}
\bibliography{references} 





\bsp	
\label{lastpage}
\end{document}

%% file: flow_chart.tex
\begin{tikzpicture}
\tikzstyle{myarrow}=[line width=0.5mm,draw=black,-triangle 45,postaction={draw, line width=0.5mm, shorten >=4mm, -}]

\node    (simulations)    {MG simulation data};
\node    (cat_true)    [below left = 0.5cm and -0.5cm of simulations]   [align=center]{halo catalogue \\ ($M_{\rm true}$)};
\node    (c_m)    [below = 0.75cm of cat_true] [align=center]   {$c_{\rm 200}(M_{\rm 500})$};
\node   (m300_m500)    [below = 1.25cm of c_m] [align=center]    {$\frac{M_{\rm 300m}}{M_{500}}$};
\node    (nfw)    [below left = 0.65cm and 1.0cm of c_m]  [anchor=west]   {NFW};
\node    (hmf_m300)    [below left = 2.7cm and 2.0cm of c_m]  [anchor=west]   {$\left( \frac{{\rm d} n_{\rm halo}}{{\rm d}M_{\rm 300m}} \right)_{f(R)}$};
\node    (hmf_th)    [below = 1.7cm of m300_m500]  [align=center]   {$\left( \frac{{\rm d} n_{\rm halo}}{{\rm d} M_{500}} \right)_{f(R)}$};
\node    (rho_eff)    [below right = 0.5cm and -0.5cm of simulations]   [align=center]  {effective density};
\node    (cat_mdyn)    [below = 1.0cm of rho_eff]    [align=center] [align=center]{halo catalogue \\ ($M_{\rm dyn}$)};
\node    (mdyn_mtrue)    [below = 1.0cm of cat_mdyn]  [align=center]   {$\frac{M_{\rm dyn}}{M_{\rm true}}(M_{\rm true})$};
\node    (observations)    [right = 5cm of simulations]    [align=center] {observational data};
\node    (n_Y)    [below = 2.0cm of observations]   [align=center]  {$\frac{{\rm d} n_{\rm cluster}}{{\rm d}Y_{\rm{obs}}}$};
\node    (hmf_obs)  at (hmf_th -| n_Y)  [ align=center]   {$\left(\frac{{\rm d} n_{\rm cluster}}{{\rm d} M_{\rm 500}}\right)_{f(R)}$};
\node    (scaling_relation)    [right = 1.75cm of mdyn_mtrue]    [align=center]{$Y_{\rm{obs}}^{f(R)}(M_{500})$};
\node    (scaling_relation_lcdm)    [below left= 0.5cm and -0cm of scaling_relation]    [align=center]{$Y_{\rm{obs}}^{\Lambda\rm{CDM}}(M_{500})$};

\node    (mcmc)    at ($(hmf_th)!0.5!(hmf_obs)+(0.0,0.25)$)   [align=center]  {MCMC};
\node    (constraint)  at ($(hmf_th)!0.5!(hmf_obs)+(0.0,-1.0)$) [align=center]  {$|f_{R0}|$ constraint};

\draw[->, line width=0.5mm] (simulations) -- (cat_true);
\draw[->, line width=0.5mm] (cat_true) -- (c_m);
\draw[->, line width=0.5mm] (c_m) -- (m300_m500);
\draw[->, line width=0.5mm, to path={-| (\tikztotarget)}] (nfw) edge (m300_m500);
\draw[->, line width=0.5mm] (m300_m500) -- (hmf_th);
\draw[->, line width=0.5mm, to path={-| (\tikztotarget)}] (hmf_m300) edge (hmf_th);
\draw[->, line width=0.5mm] (simulations) -- (rho_eff);
\draw[->, line width=0.5mm] (rho_eff) -- (cat_mdyn);
\draw[->, line width=0.5mm] (cat_mdyn) -- (mdyn_mtrue);

\draw[->, line width=0.5mm] (observations) -- (n_Y);
\draw[->, line width=0.5mm] (n_Y) -- (hmf_obs);
\draw[->, line width=0.5mm, to path={-| (\tikztotarget)}] (scaling_relation) edge (hmf_obs);
\draw[->, line width=0.5mm] (mdyn_mtrue) -- (scaling_relation);
\draw[->, line width=0.5mm, to path={|- (\tikztotarget)}] (scaling_relation_lcdm) edge (scaling_relation);

\draw[->, line width=0.5mm, to path={-| (\tikztotarget)}] (hmf_th) edge (constraint);
\draw[->, line width=0.5mm, to path={-| (\tikztotarget)}] (hmf_obs) edge (constraint);


\draw [line width=0.5mm,dotted, red, rounded corners=15pt]     ($(rho_eff.north west)+(-0.4,0.15)$) rectangle ($(mdyn_mtrue.south east)+(0.45,-0.1)$);
\node [above right = 0.10cm and -1.3cm of rho_eff] {\small{\color{red}\hypersetup{citecolor=red}\cite{Mitchell:2018qrg}}}; 
\draw[line width=0.5mm,dotted, blue, rounded corners=15pt]   ($(cat_true.north west)+(-0.4,0.15)$) rectangle ($(c_m.south east)+(0.4,-0.1)$);
\node [above left = 0.10cm and -1.3cm of cat_true] {\small{\color{blue}\hypersetup{citecolor=blue}\cite{Mitchell:2019qke}}}; 
\draw[line width=0.5mm,dotted, brown, rounded corners=15pt]     ($(hmf_th.north west)+(-0.4,0.2)$) rectangle ($(constraint.south east -| hmf_obs.south east) +(0.45,-0.1)$);
\node [above right = 0.20cm and -0.9cm of hmf_obs] {\small{\color{brown}future work}}; 
\draw[line width=0.5mm,dotted, black!60!green, rounded corners=15pt]     ($(scaling_relation.north west)+(-0.4,0.1)$) rectangle ($(scaling_relation.south east) +(0.4,-0.1)$);
\node [above right = 0.10cm and -1.3cm of scaling_relation] {\small{\color{black!60!green}this paper}}; 
\end{tikzpicture}